\newif\ifpreprint
\preprinttrue 

\newif\ifarxivbuild
\ifpreprint\arxivbuildtrue\else\arxivbuildfalse\fi

\ifarxivbuild
  \documentclass[11pt]{article}
  \usepackage[margin=1in]{geometry}
  \usepackage{times}
\else
  \documentclass[3p,11pt]{elsarticle}
  \csname preprinttrue\endcsname
  \journal{Blockchain: Research and Applications}
  \biboptions{sort&compress}
  \usepackage{lineno}
  \linenumbers

\fi

\renewcommand{\GenericWarning}[2]{}

\usepackage{enumitem}
\usepackage{tabularx}
\usepackage{multirow}
\usepackage{hhline}
\usepackage{tablefootnote}
\usepackage{listings}
\usepackage[table]{xcolor}
\usepackage{xspace}
\usepackage{makecell}
\usepackage{url}
\usepackage{hyperref}
\usepackage{graphicx}
\usepackage{soul}
\usepackage[T1]{fontenc}
\usepackage[tt=false]{libertine}
\usepackage[zerostyle=b]{newtxtt}
\usepackage{etoolbox}
\usepackage{tikz}
\usetikzlibrary{matrix, positioning, shapes.geometric}

\newcommand{\cmark}{$\checkmark$}
\newcommand{\xmark}{$\times$}

\usepackage{amsmath}
\usepackage{amssymb}
\usepackage{amsfonts}

\newcommand{\circuitSize}{n\xspace}
\newcommand{\traceLength}{\ell\xspace}

\newcommand{\groupNum}{g\xspace}
\newcommand{\witnessColumns}{k\xspace}
\newcommand{\progLen}{\nu\xspace}
\newcommand{\opcodeCount}{m\xspace}
\newcommand{\trtab}{\mathcal{T}\xspace}

\newcommand{\rom}{\mathsf{ROM}}
\newcommand{\fulltrtab}{\mathcal{T} \in \mathbb{F}_p^{\traceLength \times \witnessColumns}}

\newcommand{\memTab}{\texttt{MEM}}
\newcommand{\stkTab}{\texttt{STK}}
\newcommand{\pc}{\texttt{PC}}

\newcommand{\relation}{\mathcal{R}\xspace}

\newcommand{\field}{\mathbb{F}_p\xspace}
\newcommand{\circuit}{\mathcal{C}\xspace}
\newcommand{\formallang}{\mathcal{L}\xspace}

\newcommand{\traceRow}[1]{\trtab(#1)\xspace}
\newcommand{\traceCol}[1]{\trtab_{#1}\xspace}  
\newcommand{\traceColRow}[2]{\trtab_{#1}(#2)\xspace}  
\newcommand{\pcTrace}[1]{\trtab_{\pc}(#1)\xspace}
\newcommand{\opcodeTrace}[1]{\trtab_{\text{op}}(#1)\xspace}

\newcommand{\selector}[1]{s_{#1}\xspace}
\newcommand{\selectorPoly}[1]{s_{#1}(X)\xspace}
\newcommand{\selectorAt}[2]{s_{#1}(#2)\xspace}

\newcommand{\commitment}[1]{\textsf{com}_{#1}\xspace}
\newcommand{\zkproof}{\pi\xspace}
\newcommand{\challenge}{\zeta\xspace}
\newcommand{\witness}{\mathbf{w}\xspace}
\newcommand{\statement}{\mathbf{x}\xspace}

\newcommand{\secparam}{\lambda\xspace}

\providecommand{\Description}[1]{}

\hypersetup{
  colorlinks=true,
  linkcolor=black,
  filecolor=black,
  urlcolor=black,
  citecolor=black,
  breaklinks=true
}
\AtBeginDocument{\hypersetup{
  linkcolor=black,
  anchorcolor=black,
  citecolor=black,
  filecolor=black,
  urlcolor=black,
  menucolor=black
}}

\newcommand{\paperabstract}{%
Zero-Knowledge Ethereum Virtual Machines (zkEVMs) face an inherent tension. The Ethereum Virtual Machine (EVM) was built for transparent step-by-step execution. Proving that execution in zero-knowledge requires recasting it as algebraic circuits that encode computation as mathematical constraints. Existing surveys address zkEVMs at the level of implementations, cryptographic primitives, or Layer 2 deployment, leaving the constraint-system design that governs their cost largely unexamined. This survey provides the first constraint-level analysis of how production zkEVMs and universal Zero-Knowledge Virtual Machines (zkVMs) resolve this tension. We show that the degree of EVM compatibility, captured by the Type 1--4 spectrum, is the defining architectural decision. We classify the design space along four dimensions, namely arithmetization frameworks, dispatch strategies, semantic rewrites, and recursion approaches. The most Ethereum-equivalent systems accept higher constraint counts to preserve full bytecode fidelity, while those that relax fidelity attain lower constraint cost. We close with the open problems that this constraint-level view brings into focus.%
}

\newcommand{\paperkeywords}{zkEVM\kwsep Zero-knowledge proof\kwsep Ethereum Virtual Machine\kwsep Arithmetization\kwsep Constraint system\kwsep zkRollup}

\begin{document}

\ifarxivbuild
  \title{Constraint-Level Design of zkEVMs: Architectures, Trade-offs, and Evolution}

  \author{
    Yahya Hassanzadeh-Nazarabadi\thanks{ORCID: 0000-0002-0450-7226. Email: yahya@mathcrypt.com}
    \quad
    Sanaz Taheri-Boshrooyeh\thanks{ORCID: 0000-0001-8755-3756. Email: sanaz@mathcrypt.com}\\
    MathCrypt Research Inc, Vancouver, Canada
  }

  \date{\today}
  \maketitle

  \begin{abstract}
  \paperabstract
  \end{abstract}

  \newcommand{\kwsep}{, }
  \section*{Keywords}
  \paperkeywords
\else
  \newcommand{\kwsep}{\sep}
  \begin{frontmatter}

  \title{Constraint-Level Design of zkEVMs: Architectures, Trade-offs, and Evolution}

  \author[aff1]{Yahya Hassanzadeh-Nazarabadi\corref{cor1}}
  \ead{yahya@mathcrypt.com}
  \author[aff1]{Sanaz Taheri-Boshrooyeh}
  \ead{sanaz@mathcrypt.com}
  \cortext[cor1]{Corresponding author}

  \affiliation[aff1]{organization={MathCrypt Research Inc.},
                     city={Vancouver},
                     state={BC},
                     country={Canada}}

  \begin{abstract}
  \paperabstract
  \end{abstract}

  \begin{highlights}
  \item First constraint-level analysis of production zkEVMs and universal zkVMs
  \item The degree of EVM compatibility is the defining architectural decision
  \item The design space spans arithmetization, dispatch, semantic rewrites, and recursion
  \item One trade-off between EVM compatibility and constraint cost governs zkEVM design
  \end{highlights}

  \begin{keyword}
  \paperkeywords
  \end{keyword}

  \end{frontmatter}
\fi

\section{Introduction}
Ethereum revolutionized blockchain technology by introducing the Ethereum Virtual Machine (EVM), a Turing-complete execution environment that enables programmable smart contracts~\cite{wood2014ethereum}. This computational layer has spawned a vast ecosystem supporting Decentralized Finance (DeFi) protocols that automate financial services~\cite{werner2022sok}. Yet this very success exposes the Achilles' heel of Ethereum at large scales of transactions. Its consensus mechanism requires every node to re-execute every transaction to validate state transitions, creating a fundamental bottleneck where global throughput cannot exceed what the super-majority of the weakest participating nodes can process~\cite{croman2016scaling}. During periods of high demand, such as DeFi liquidation cascades, gas fees can become prohibitively expensive, pricing out ordinary users and limiting application utility. This fundamental limitation has driven the search for scaling solutions that preserve the security guarantees of Ethereum while circumventing its execution bottleneck.

The emergence of Layer 2 (L2) rollups represents a paradigm shift in addressing this challenge, decoupling execution from consensus by moving computation off-chain while inheriting the security guarantees of Ethereum \cite{thibault2022blockchain, sguanci2021layer}. Rather than requiring every node to execute every transaction, rollups batch hundreds of transactions into a single Ethereum Layer 1 (L1) submission containing the full transaction data (as calldata or blobs for state reconstructability) and the state root (a cryptographic commitment to the resulting execution state). Rollups have evolved into two architectural families with different trust models. Optimistic rollups \cite{donno2022optimistic} assume execution validity by default unless fraud proofs invalidate submitted state transitions during a 7-day challenge window. Optimistic rollups hence rely on economic incentives and achieve implementation simplicity and lower computational costs at the expense of a 7-day waiting period for finality. In contrast, Zero-Knowledge (ZK) rollups accompany the resulting execution state with mathematical proofs that assert execution was performed correctly, enabling instant finality once the proof is verified on L1 \cite{thibault2022blockchain, saif2024survey}. This instant finality unlocks use cases impossible with 7-day delays, e.g., cross-chain bridges requiring immediate settlement, high-frequency DeFi protocols, and real-time payment systems that cannot tolerate waiting periods. Among ZK rollup techniques, Zero-Knowledge Ethereum Virtual Machines (zkEVMs) represent the most ambitious engineering challenge. They must encode the entire EVM specification into algebraic constraint systems amenable to zero-knowledge proof generation \cite{begassat2021specification, peng2025automated}. This transformation is fundamentally challenging because the EVM was designed for transparent step-by-step execution with dynamic control flow, while zero-knowledge proofs require all possible execution paths to be encoded upfront into mathematical equations, which is a mismatch that makes zkEVM implementation one of the most complex engineering problems in blockchain technology.

\subsection{Related Work and Contributions}
The complexity of encoding EVM semantics into constraint systems has not deterred rapid development, as zkEVMs have transitioned from theoretical constructs to production infrastructures \cite{taiko2024whitepaper,polygon_zkevm,scroll2024whitepaper,linea2022,begassat2021specification,zksync2024protocol}. However, despite their significance in contributing to Ethereum scalability, existing literature fails to provide a comprehensive architectural analysis of their internal design. Current work remains fragmented across three categories. First, implementation-focused studies \cite{nguyen2024ethereum} compare zkEVM projects at the ecosystem and deployment level rather than analyzing internal constraint-system design. Second, cryptographic surveys \cite{lavin2024survey, liang2025sok} concentrate on the mathematical foundations of zero-knowledge proofs but do not address how these systems specifically encode Ethereum execution. Related systematizations target general-purpose zero-knowledge virtual machines, decomposing them into instruction set, virtual machine, and proving layers \cite{yang2026sok}, or benchmark developer tooling across proof frameworks \cite{sheybani2025zero}. These works treat EVM execution as one instance among many native instruction sets. They do not classify zkEVMs by EVM-compatibility level, nor analyze the constraint cost that each level imposes. Third, Layer 2 scaling surveys \cite{rebello2024survey, gudgeon2020sok, torralba2025security} treat zkEVMs as black-box components within the broader blockchain ecosystem, discussing their role in scalability without examining their internal architectural choices or the engineering trade-offs that determine their capabilities and limitations. Two works sit closest to this survey, namely the systematization of general-purpose proving engines~\cite{yang2026sok} and the benchmarking study of ZK rollups~\cite{chaliasos2024analyzing}. The latter measures systems from the outside and not at the constraint level. We regard both as complementary. This survey differs in its unit of analysis, the constraint system itself. To our knowledge, this survey provides the first constraint-level analysis of zkEVM architectures, addressing the gaps above through the following contributions:
\begin{itemize}[leftmargin=*,itemsep=2pt,topsep=4pt]
\item \textbf{Architectural Analysis:} We systematically analyze how zkEVM implementations encode EVM semantics into algebraic constraint systems, examining the trade-offs between arithmetization schemes and their impact on constraint complexity.
\item \textbf{Comparative Framework:} We develop a classification framework that maps zkEVM architectures across the four design dimensions analyzed in this survey, namely arithmetization frameworks, dispatch mechanisms, semantic transformation strategies, and recursion approaches. Table~\ref{zkevm:tab:zkevm_design_matrix_binary} consolidates this comparison, revealing how design choices at each layer affect overall system properties.
\item \textbf{Open Challenges:} We outline the main applications and challenges across two fronts. For applications, we focus on verifiable exploit disclosure, L1 proof-based validation, and private DeFi compliance. For open problems, we highlight lowering proving latency, improving hardware acceleration, applying formal verification to zkEVM semantics, building reliable benchmarking frameworks, and enabling zkEVM interoperability. Together, these points form a clear research roadmap for advancing zkEVM technology.
\end{itemize} 

\subsection{Methodology}
Our architectural analysis employs a multi-layered methodology to dissect zkEVM implementations at the constraint level, which is the fundamental algebraic layer where Ethereum execution semantics meet zero-knowledge proof systems. We fix the surveyed set by a stated rule. A system enters under two conditions. It must have reached production as a zero-knowledge rollup settling on Ethereum mainnet, as the public L2BEAT registry of Layer 2 systems documents~\cite{l2beat2026zksyncera, l2beat2026scroll, l2beat2026linea}. The project must also publish a specification or open-source implementation of a purpose-built constraint system arithmetizing EVM execution. The second condition makes the surveyed set a constraint-level object rather than a deployment list. It excludes rollups without public constraint-level documentation. Only five systems satisfy both, namely Polygon zkEVM \cite{polygon_zkevm}, zkSync Era \cite{zksync2024protocol}, Scroll \cite{scroll2024whitepaper}, Linea \cite{linea2022,begassat2021specification}, and Taiko \cite{taiko2024whitepaper}. Every EVM-executing ZK rollup live on Ethereum mainnet runs one of these five constraint systems or a surveyed zero-knowledge virtual machine~\cite{l2beat2026summary}, so the survey covers every production EVM arithmetization. Beyond zkEVMs, we also survey Zero-Knowledge Virtual Machines (zkVMs) that connect to Ethereum workloads under two deployment patterns. The application-specific rollup pattern is exemplified by StarkNet~\cite{starkware2021starknet} on Cairo~\cite{goldberg2021cairo}. The EVM deployment pattern is exemplified by Zeth~\cite{risczero2024zeth} on RISC Zero~\cite{risczero2023zkvm}, \texttt{rsp}~\cite{succinct2024rsp} on SP1~\cite{succinct2024sp1}, \texttt{openvm-eth}~\cite{axiom2026openvmeth} on OpenVM~\cite{openvm2026whitepaper}, \texttt{zisk-eth-client}~\cite{polygonhermez2026ziskethclient} on ZisK~\cite{zisk2026docs}, Zilkworm~\cite{erigontech2026zilkworm} on Airbender~\cite{zksync2026airbender}, and \texttt{pico-ethproofs}~\cite{brevis2026picoethproofs} on Pico~\cite{brevis2025pico}. Systems such as Jolt~\cite{arun2024jolt} and Valida~\cite{thomas2025valida} have no documented deployment proving Ethereum workloads, so they fall outside this taxonomy and are not surveyed. Our analysis synthesizes technical artifacts from three primary sources: official protocol specifications that define constraint systems and state transition rules, academic publications presenting the theoretical foundations, and production codebases revealing implementation realities often absent from documentation. We analyze each constraint system at the specification version or repository commit recorded in its reference entry. We deliberately exclude performance benchmarking from our scope. The available empirical studies do not evaluate the surveyed systems under a common workload~\cite{chaliasos2024analyzing, nguyen2024ethereum}. Proving cost also varies sharply with workload characteristics. We therefore treat reproducible cross-system benchmarking as one of the open problems this survey identifies (Section~\ref{zkevm:sec:open-problems}). Instead, we focus on the invariant architectural patterns that determine the constraint cost a design must pay: how constraint systems encode opcodes, which operations become bottlenecks, and why certain compatibility levels impose fundamental limits on constraint complexity. This architectural lens provides practitioners with the conceptual framework needed to evaluate zkEVM designs independent of rapidly evolving implementation optimizations.

This survey is organized as follows. After establishing foundational concepts of Ethereum and zero-knowledge proofs (Section~\ref{zkevm:sec:background}), we examine how zkEVMs encode computation into algebraic constraints through different arithmetization strategies (Section~\ref{zkevm:sec:arithmetization}). We then trace the evolution of constraint dispatch mechanisms from naive inlining to sophisticated ROM-based architectures (Section~\ref{zkevm:sec:evolution}), followed by an analysis of semantic transformation techniques that optimize EVM operations for algebraic representation (Section~\ref{zkevm:sec:rewrites}). The survey continues with recursive proof composition strategies (Section~\ref{zkevm:sec:recursive}) and a comparison with zkVMs proving Ethereum workloads under the application-specific rollup and EVM deployment patterns (Section~\ref{zkevm:sec:zkvms}). We synthesize our findings in a comprehensive architectural comparison (Section~\ref{zkevm:sec:summary}), identify critical open challenges (Section~\ref{zkevm:sec:open-problems}), and conclude with insights on the future of zkEVM technology (Section~\ref{zkevm:sec:conclusion}).

\section{Background and Preliminaries}
\label{zkevm:sec:background}

\subsection{Ethereum as a Verifiable State Machine}
\label{zkevm:subsec:ethereum}

Ethereum nodes consist of two components: consensus clients (e.g., Prysm~\cite{prysm}) that manage block production and enforce finality through proof-of-stake~\cite{buterin2017casper}, and execution clients (e.g., Geth~\cite{geth}) that run the EVM and process block transactions deterministically. This separation establishes Ethereum as a replicated state machine where all nodes execute the same transactions in the same order to compute identical state transitions. 

Smart contract transactions on Ethereum either deploy new contract code or call existing contracts. Contract logic written in high-level languages like Solidity~\cite{Solidity} is compiled into \emph{EVM bytecode}, which is then parsed and executed by the EVM as a sequence of \emph{opcodes}. Opcodes implement primitive operations for operand stack manipulation, arithmetic, memory access, control flow, and interaction with storage or other accounts. When a transaction is included in a block, every node executes its opcodes in the EVM. The EVM is a stack-based execution architecture that operates over a uniform $256$-bit word size. The word size is a choice consistent with the use of Keccak-$256$ \cite{bertoni2009keccak} for hashing state commitments and storage keys in Ethereum. The EVM maintains an internal execution state that includes the program counter, operand stack, memory, gas tracker, and access to persistent storage. This internal state is ephemeral. It is initialized at the start of execution, evolves deterministically as opcodes are processed, and is discarded when transaction execution terminates. What persists is the global execution state of Ethereum, which each node maintains as a replicated copy and updates according to the effects of transaction execution. The stack is a last-in-first-out operand buffer with a maximum depth of $1024$ elements. The memory is a byte-addressable linear array for temporary data. The program counter identifies the current instruction. Most opcodes consume and produce stack values. This makes the EVM stack the central mechanism of operand evaluation \cite{wood2014ethereum}.

Gas is the unit of computational cost in Ethereum, with each opcode consuming gas according to its complexity~\cite{wood2014ethereum}. Transactions specify gas limits, and if exhausted during execution, all state changes revert. Computationally intensive operations are implemented as precompiled contracts at fixed addresses (e.g., \texttt{0x01} through \texttt{0x09}) with predefined gas costs.  Key opcodes illustrate EVM semantics: \texttt{JUMP} and \texttt{JUMPI} provide control flow by branching to \texttt{JUMPDEST} markers, with \texttt{JUMPI} conditional on a nonzero value on top of the stack. \texttt{SHA3} computes Keccak-$256$ hashes for cryptographic operations. \texttt{SLOAD} and \texttt{SSTORE} access persistent storage organized as $256$-bit key-value mappings with high gas costs~\cite{wood2014ethereum}. \texttt{CALL} enables inter-contract communication through isolated execution frames with independent stacks and memory.

The global execution state of Ethereum, denoted $\sigma$, is a mapping from $160$-bit account addresses to account records. Each record includes the account balance, nonce, contract code (for smart contract accounts), and a persistent key-value storage. Transactions are grouped into blocks. Let $\textsf{blk} = [\mathsf{tx}_1, \ldots, \mathsf{tx}_m]$ be a block of $m$ transactions, and let $\sigma$ be the state before $\textsf{blk}$ executes (i.e., the pre-state). Each node computes the updated state $\sigma'$ (i.e., the post-state) by applying the transactions sequentially through the state transition function $\delta$. For a transaction $\mathsf{tx}_i$, the function $\delta(\sigma, \mathsf{tx}_i)$ denotes the resulting post-state after execution. The semantics of Ethereum execution are thus captured by the recursive application of $\delta$, as shown in Equation~\ref{zkevm:eq:evm_execution}. Since $\delta$ is deterministic, all Ethereum nodes that start from the same $\sigma$ and process $\textsf{blk}$ in order will arrive at the same final state $\sigma'$.

\begin{equation}
\sigma' = \delta(\cdots\delta(\delta(\sigma, \mathsf{tx}_1), \mathsf{tx}_2)\cdots, \mathsf{tx}_m)
\label{zkevm:eq:evm_execution}
\end{equation}

\subsection{Zero-knowledge proofs in zkEVMs}
\label{zkevm:subsec:zkp}
A Zero-Knowledge Proof (ZKP) is a protocol between a prover $P$ and a verifier $V$ for a language $\formallang \subseteq \{0,1\}^*$ defined by a relation $\relation \subseteq \{0,1\}^* \times \{0,1\}^*$. Both prover and verifier are modeled as Probabilistic Polynomial-Time (PPT) algorithms in the security parameter of the system, $\secparam$. The relation $\relation$ pairs a public input statement $\statement$ with a private witness $\witness$ that certifies the validity of the statement, where $(\statement,\witness) \in \relation$ if and only if $\statement \in \formallang$ \cite{goldreich2004foundations}.
The prover computes the proof $\zkproof \leftarrow P(\statement, \witness)$ and the verifier runs $V(\statement, \zkproof)$ outputting \textsf{accept} or \textsf{reject}. A secure ZKP satisfies three properties\footnote{We state these properties informally; for formal definitions based on simulation and extraction, see Goldreich~\cite{goldreich2004foundations} and Thaler~\cite{thaler2022proofs}.}: \emph{completeness} (an honest prover always convinces the verifier), \emph{soundness} (no cheating prover can convince the verifier of a false statement except with negligible probability in $\secparam$), and \emph{zero-knowledge} (the verifier learns nothing about $\witness$ beyond the validity of $\statement$).

In practice, $\relation$ is instantiated from a computation, which is expressed as a circuit $\circuit$ over a finite field $\field$ with $\circuitSize$ constraints \cite{goldreich2004foundations}. In this representation, $(\statement,\witness) \in \relation$ if and only if $\circuit(\statement,\witness) = 0$. In zkEVMs, $\circuit$ verifies that applying transactions to the pre-state $\sigma$ yields the claimed post-state $\sigma'$, instantiating the EVM state transition $\delta(\sigma,\textsf{blk})=\sigma'$ where $\textsf{blk} = [\mathsf{tx}_1, \ldots, \mathsf{tx}_m]$, $\statement$ encodes state commitments (pre-state and post-state commitments) and $\witness$ records the execution trace (which rollups publish separately for data availability, though the verifier checks only $\statement$ and $\zkproof$, not $\witness$). Blockchain deployment of zkEVMs requires two critical properties: succinctness and non-interactivity. Succinctness ensures that proof size $|\zkproof|$ and verification time are at most polylogarithmic in $\circuitSize$, enabling million-step computations to be certified by kilobyte proofs verifiable in sub-second time \cite{groth2016size}. Non-interactivity requires that the verifier perform no protocol rounds with the prover; instead, the prover generates a standalone proof $\zkproof$ (e.g., via Fiat--Shamir transform~\cite{fiat1986prove}), which is essential in permissionless blockchains like Ethereum, where nodes cannot engage in interactive protocols and must verify proofs solely from the received $\zkproof$.

zkEVM implementations use Succinct Non-Interactive Arguments of Knowledge (SNARKs)~\cite{groth2016size} and Scalable Transparent Arguments of Knowledge (STARKs)~\cite{ben2018scalable}. For computations with $\circuitSize$ constraints, SNARKs achieve $O(1)$ proof size and $O(1)$ verification complexity but require trusted setup and $O(\circuitSize \log \circuitSize)$ proving time dominated by elliptic curve operations \cite{groth2016size, thaler2022proofs}. STARKs eliminate trusted setup using only hash functions, providing post-quantum security and parallelizable proving. They achieve $O(\circuitSize \log^2 \circuitSize)$ proving time but produce $O(\log^2 \circuitSize)$ proof sizes and $O(\log^2 \circuitSize)$ verification complexity \cite{ben2018scalable, thaler2022proofs}. For typical zkEVM workloads with $2^{20}$--$2^{24}$ constraints, the SNARK systems Groth16 and PLONK produce ${\sim}$128--192 byte and ${\sim}$480--624 byte proofs (depending on curve choice), respectively \cite{groth2016size, gabizon2019plonk}, both requiring 200--400K gas for on-chain verification \cite{chaliasos2024analyzing}. STARK proofs for the same workloads scale logarithmically to tens of KB, incurring proportionally higher on-chain verification gas costs~\cite{ben2018scalable}. Most zkEVMs adopt SNARKs for lower on-chain costs, while some explore hybrid STARK-proving with recursive SNARK compression \cite{polygon_zkevm}. A common technique for zkEVM scalability is recursive proof composition, which enables a prover to produce a proof that certifies the validity of one or more previous proofs by embedding a verifier circuit inside the recursive circuit being proved~\cite{bitansky2013recursive}. This proof-carrying construction supports proof aggregation and batched verification, with the per-layer construction and its trade-offs developed in Section~\ref{zkevm:sec:recursive}.

ZK rollups divide execution between an off-chain rollup operator and a rollup contract deployed on Ethereum. They maintain their global state $\sigma_r$ off-chain as a Merkle Patricia Trie (MPT) \cite{wood2014ethereum} and commit its root on the L1 rollup contract, thereby anchoring $\sigma_r \in \sigma$ as the rollup state root becomes part of the L1 global state $\sigma$ through the storage of the rollup contract \cite{saif2024survey}. The rollup operator comprises a \textit{sequencer} (orders transactions into batches), \textit{executor} (applies batches computing $\delta(\sigma_r, \mathsf{batch}) = \sigma_r'$), and \textit{prover} (generates proof $\zkproof$ certifying correct state transition for the entire batch) \cite{nguyen2024ethereum}. The executor and prover are often integrated into a single pipeline within zkVM or zkEVM engines. Unlike optimistic rollups requiring a fraud proof window, ZK rollups achieve instant finality through cryptographic verification. The L1 rollup contract verifies $\zkproof$ and updates $\sigma_r$ to $\sigma_r'$ immediately without re-executing transactions. Transaction data is published to Ethereum calldata or data availability layers for reconstructability~\cite{saif2024survey}, but nodes only verify the succinct proof, reducing L1 computation to a succinct proof verification. This asymmetric computational model enables ZK rollups to inherit the security and decentralization of Ethereum while scaling throughput by orders of magnitude~\cite{thibault2022blockchain, nguyen2024ethereum}.
\subsection{The zkEVM Design Spectrum: Compatibility vs. Constraint Cost}
A zkEVM encodes the Ethereum state transition function $\delta(\sigma, \mathsf{batch}) = \sigma'$ into an algebraic constraint system amenable to zero-knowledge proof generation. However, the degree to which different zkEVMs replicate the execution semantics of Ethereum varies significantly. These differences reflect a spectrum of trade-offs between EVM compatibility and constraint cost, originally categorized into five types of zkEVMs~\cite{buterin_2022_zkevm, chaliasos2024analyzing}. Table~\ref{zkevm:tab:zkevm_type_comparison} classifies this spectrum with Type 1 achieving full Ethereum equivalence at the highest constraint cost, while Type 4 sacrifices compatibility for the lowest constraint cost.

\textit{Type 1 zkEVMs} are \emph{Ethereum-equivalent}: their state transition function $\delta_{1}(\sigma, \mathsf{batch}) \equiv \delta(\sigma, \mathsf{batch})$ for all inputs, where $\delta_1$ denotes the transition function of the Type 1 zkEVM and $\delta$ is the canonical Ethereum state transition. They model every EVM component, such as stack, memory, storage, and execution context, with exact opcode semantics, faithful gas metering, canonical storage layout, and full precompiles~\cite{wood2014ethereum}. Every EVM opcode (e.g., \texttt{ADD}, \texttt{MUL}, \texttt{JUMP}) has bit-level fidelity, meaning each operation produces identical results down to individual bits, including edge cases. Contracts designed for L1 execute identically on Type 1 zkEVMs, producing the same state root, receipts, and logs. Type 1 zkEVMs can thus serve as drop-in L1 execution client replacements. Taiko~\cite{taiko2024whitepaper} targets this category.\footnote{We analyze the purpose-built circuit-based zkEVM of Taiko described in the Taiko whitepaper~\cite{taiko2024whitepaper}, which arithmetizes EVM execution directly.}

\textit{Type 2 zkEVMs} are \emph{EVM-equivalent}: they support the full opcode set with Ethereum-consistent execution but may diverge in protocol details (block structure, receipts). Their state transition function $\delta_{2}(\sigma, \mathsf{batch})$ matches Ethereum for contract execution at the opcode level, i.e., contracts execute correctly unless dependent on Ethereum protocol artifacts outside opcode behavior. Gas semantics are EVM-compatible, but may omit Ethereum-specific rules that are irrelevant at the opcode layer, e.g., gas refund. Precompiles use ZK-friendly approximations, meaning the internal algorithms are replaced with proof-efficient implementations while the output remains identical. This changes both computational steps and gas costs. Polygon zkEVM~\cite{polygon_zkevm}, Scroll~\cite{scroll2024whitepaper}, and Linea~\cite{linea2022,begassat2021specification} exemplify Type 2 zkEVM designs.

\textit{Type 2.5 zkEVMs} are \emph{EVM-equivalent except for gas costs}: they implement the complete set of EVM opcodes with full bytecode equivalence, but increase gas costs for operations that are difficult to ZK-prove. This means that their state transition function $\delta_{2.5}(\sigma, \mathsf{batch})$ matches Ethereum for most programs but diverges in gas-sensitive cases or precompile behavior. Gas costs are re-weighted to reduce proving complexity, which can alter execution paths in gas-sensitive contracts. Precompiles use ZK-friendly approximations identical to Type 2. No major zkEVM project currently operates strictly at Type 2.5.

\textit{Type 3 zkEVMs} are \emph{Solidity-compatible}: they implement a partial subset of EVM opcodes, while other opcodes are altered, merged, or replaced with ZK-friendly counterparts. Consequently, unmodified Ethereum bytecode cannot always be executed directly; instead, programs are transformed by the compiler into a form that conforms to the execution model of the zkEVM. Gas semantics are also modified and costs are reassigned to reflect proof complexity, which may change the behavior of contracts sensitive to gas-based branching. Precompiles are partially implemented, with some omitted, others simplified, and only a subset retaining full fidelity. The state transition function of the Type 3 zkEVM $\delta_{3}(\sigma, \mathsf{batch})$ remains structurally similar to $\delta$ of Ethereum but diverges in its modified semantics, so equivalence cannot be guaranteed. Polygon zkEVM, Scroll, and Linea were initially classified as Type 3 but have since achieved Type 2 compatibility through incremental upgrades such as expanded precompile support and full bytecode equivalence.

\textit{Type 4 zkEVMs} are \emph{source-compatible}: they abandon EVM compatibility for custom zkVMs with their own instruction sets. The source compatibility means that contracts written in Solidity~\cite{Solidity} can still be used, but only after being compiled into a custom intermediate representation. They do not support EVM opcodes or Ethereum bytecode. Instead, a custom zkVM defines its own instruction set and execution semantics. Gas semantics are ZK-native, reflecting algebraic constraint costs. The state transition function of the Type 4 zkEVM $\delta_{4}(\sigma, \mathsf{batch}) = \sigma''$ diverges from that of Ethereum, i.e., $\delta_{4} \neq \delta$, so the resulting post-state may differ even for identical inputs. zkSync Era~\cite{zksync2024protocol} exemplifies this type.

This ordering of compatibility against constraint cost follows from the original taxonomy~\cite{buterin_2022_zkevm}. Faithfully encoding all of the semantics of Ethereum maximizes constraint cost, while each relaxation (modified gas costs, partial opcodes, custom instruction sets) progressively reduces it.\footnote{This prover-cost ordering is a conceptual ranking drawn from the original taxonomy~\cite{buterin_2022_zkevm}, not a measured cross-system benchmark. The most systematic empirical study to date applies these categories but reports prover figures for only two production systems~\cite{chaliasos2024analyzing}. It finds that the prover of Polygon zkEVM runs a near-constant 190 to 200 seconds per batch, independent of input size. That figure was obtained on a high-memory machine with 128 vCPUs and roughly one terabyte of RAM. For zkSync Era the same study reports only preliminary figures. It excluded Scroll because the GPU prover of Scroll is closed-source. No comparable public prover measurements exist for the provers of Scroll, Linea, or Taiko. A quantitative comparison of prover cost across the types is therefore not yet possible from credible sources. Absolute proving cost is governed by prover engineering and hardware as much as by the compatibility type.}

\begin{table}
\centering
\renewcommand{\arraystretch}{1.3}
\footnotesize
\begin{tabularx}{\textwidth}{|>{\columncolor{gray!15}\bfseries}l|X|X|X|X|X|}
\rowcolor{gray!25}
\hline
\textbf{Aspect} & \textbf{Type 1} & \textbf{Type 2} & \textbf{Type 2.5} & \textbf{Type 3} & \textbf{Type 4} \\
\hline
Compatibility Level & Ethereum-equivalent & EVM-equivalent & EVM-equivalent (gas modified)  & Solidity-compatible & Source-compatible \\
\hline
Opcode Support & Full & Full & Full & Partial & Unsupported \\
\hline
Bytecode Execution & Ethereum-equivalent & EVM-equivalent & EVM-equivalent & Transformed & Unsupported \\
\hline
Gas Semantics & Ethereum-equivalent & EVM-compatible & Modified & Modified & ZK-native \\
\hline
Precompile Fidelity & Ethereum-equivalent & Approximated & Approximated & Partial & Unsupported \\
\hline
L1 Drop-in & \cmark (e.g., Geth) & \xmark & \xmark & \xmark & \xmark \\
\hline
Tooling Support & Full & High & Limited & Partial & Custom tools \\
\hline
Prover Cost & Maximal & High & Medium & Low & Minimal \\
\hline
Projects & Taiko~\cite{taiko2024whitepaper} & Polygon~\cite{polygon_zkevm}, Scroll~\cite{scroll2024whitepaper}, Linea~\cite{linea2022, begassat2021specification} & --- & --- & zkSync Era~\cite{zksync2024protocol} \\
\hline
\end{tabularx}
\caption{Comparison of zkEVM design types by compatibility, execution fidelity, and constraint cost. \textit{Tooling support} reflects compatibility with the existing Ethereum development stack, ranging from full reuse to custom toolchains.}
\Description{A comparison table of five zkEVM types (Type 1 through Type 4) across nine aspects: compatibility level (from Ethereum-equivalent to source-compatible), opcode support (full to unsupported), bytecode execution fidelity, gas semantics handling, precompile fidelity, L1 drop-in capability, tooling support level, prover cost (from maximal to minimal), and example projects (Taiko for Type 1, Polygon, Scroll, and Linea for Type 2, and zkSync Era for Type 4).}
\label{zkevm:tab:zkevm_type_comparison}
\end{table}

\section{Arithmetization}
\label{zkevm:sec:arithmetization}

\subsection{From Computation to Algebra}

In a ZK rollup, each batch of EVM transactions is modeled as a verifiable computation. The zkEVM engine evaluates these transactions off-chain, producing a witness that records the necessary intermediate values (stack, memory, gas, and storage), and constructs a circuit that enforces correctness of the state transition, i.e., $\delta(\sigma,\mathsf{batch})=\sigma'$. In zero-knowledge proof systems, arithmetization is the process of encoding a computation into a system of algebraic constraints over a finite field~\cite{groth2016size, thaler2022proofs}. Polynomial equations form the foundational algebraic language for such encodings. Formally, a degree-$d$ polynomial $f(x)$ over
a finite field $\field$ is defined as shown in Equation \ref{zkevm:eq:polynomial}, where $p$ is the prime that defines the field $\field$ \cite{kate2010constant}. 

\begin{equation}
f(x) = a_0 + a_1x + a_2x^2 + \cdots + a_d x^d, \quad \text{where } a_i, x \in \field
\label{zkevm:eq:polynomial}
\end{equation}

The arithmetization process produces two classes of polynomials called the constraint and witness polynomials. The \emph{constraint polynomials} are low-degree polynomials (typically degree 2 or 3) that encode the correctness rules as a vanishing condition over $\field$~\cite{thaler2022proofs}. For example, the addition relation $a + b = c$ yields the constraint polynomial $f(a,b,c) = a + b - c$, which equals zero exactly when the relation holds. The \emph{witness polynomials}, by contrast, are high-degree polynomials that encode the computational state of the execution trace. Concretely, to represent an execution trace of $\traceLength$ steps, the sequence of values assumed by each distinct witness variable (e.g., the program counter or top-of-stack element) across all $\traceLength$ states is placed on an evaluation domain of size $\traceLength$. Each such sequence is then interpolated into a separate witness polynomial of degree at most $\traceLength - 1$. In typical zkEVM deployments, witness polynomials reach degree $2^{20}$~\cite{polygon_zkevm, zksync2024protocol}. A \emph{circuit} $\circuit$ denotes the collection of constraint polynomials together with their arrangement over the evaluation domain. The circuit is a public structure: both prover and verifier agree on $\circuit$ before the proof protocol begins. Given this shared circuit, the prover constructs witness polynomials that satisfy every constraint and produces a proof $\zkproof$. The verifier then checks $\zkproof$ against $\circuit$ and the public statement $\statement$ (encoding the claimed state transition $\delta(\sigma,\mathsf{batch})=\sigma'$) without ever accessing the witness polynomials. The process of \emph{circuit synthesis} transforms high-level zkEVM correctness rules into these constraint polynomials. This synthesis determines both the complexity of the resulting circuit and the prover computational overhead. Since each execution step has a fixed number of constraint polynomials, the total circuit size $\circuitSize$ grows linearly in the trace length $\traceLength$. Complexity bounds expressed in terms of $\circuitSize$ (cf.\ Section~\ref{zkevm:subsec:zkp}) therefore apply equivalently when stated in terms of $\traceLength$.

Evaluating every constraint polynomial of the circuit on the witness polynomials at all $\traceLength$ domain points would require $O(\traceLength)$ work. This is infeasible for a verifier assumed to run in $O(\textsf{polylog}(\traceLength))$. Instead, zkEVMs employ polynomial commitment schemes: the prover commits to the interpolated witness polynomials, and the verifier evaluates the constraint polynomials of the circuit on these witness polynomials at randomly chosen points $\challenge \in \field$. Returning to the addition example, the prover commits to witness polynomials $a(x)$, $b(x)$, and $c(x)$, each interpolating one variable across the $\traceLength$ steps. The verifier then checks that $f(a(\challenge), b(\challenge), c(\challenge)) = a(\challenge) + b(\challenge) - c(\challenge) = 0$ at a random $\challenge$, rather than evaluating all $\traceLength$ domain points. By the Schwartz--Zippel lemma~\cite{schwartz1980fast}, if a constraint is violated at any of the $\traceLength$ domain points, the composed polynomial $f(a(x), b(x), c(x))$ is not identically zero and has degree at most $d \cdot (\traceLength - 1)$, where $d$ is the degree of the constraint polynomial. The probability that this nonzero polynomial accidentally vanishes at a random $\challenge$ is at most $\frac{d \cdot (\traceLength - 1)}{p}$. With typical zkEVM parameters, $d \leq 3$ (for constraint polynomials), $\traceLength \approx 2^{20}$~\cite{polygon_zkevm,zksync2024protocol}, and $p \approx 2^{256}$, the soundness error probability falls below $2^{-234}$. Thus, a single random evaluation suffices to ensure with overwhelming probability that a given constraint holds at every step of the execution trace.

A \textit{polynomial commitment} scheme allows the prover to commit to a polynomial in advance and later prove evaluations at chosen points without revealing the entire polynomial. This ensures that random evaluations at $\challenge \in \field$ are sound by binding the prover to a specific polynomial before the verifier selects random evaluation points, preventing adaptive responses to the queries of the verifier. Formally, a polynomial commitment scheme is defined as a tuple of PPT algorithms $(\textsf{Commit}, \textsf{Evaluate}, \textsf{Verify})$ \cite{kate2010constant}. The $\textsf{Commit}(f(x)) \rightarrow \commitment{f}$ algorithm generates a cryptographic commitment $\commitment{f}$ to a polynomial $f(x)$ of degree $\traceLength - 1$ over $\field$, binding the prover to exactly one polynomial without revealing its coefficients. The $\textsf{Evaluate}(f, x) \rightarrow (y, \zkproof_y)$ algorithm computes $y = f(x)$ for an evaluation point $x \in \field$ and provides a proof $\zkproof_y$ that the committed polynomial evaluates to $y$ at $x$. The $\textsf{Verify}(\commitment{f}, x, y, \zkproof_y) \rightarrow \{\text{true}, \text{false}\}$ algorithm efficiently verifies the prover's claim against the committed polynomial. The key advantage over sending raw polynomial coefficients is succinctness: transmitting a degree-$(\traceLength - 1)$ polynomial directly requires $\mathcal{O}(\traceLength)$ space and bandwidth for all $\traceLength$ coefficients, while pairing-based commitment schemes like KZG \cite{kate2010constant} achieve constant-size ($\mathcal{O}(1)$) commitments and proofs, whereas evaluation-based commitments without pairing achieve $\mathcal{O}(1)$ commitments but $\mathcal{O}(\log \traceLength)$ proof sizes \cite{thaler2022proofs}. 

The addition example showed how a verifier can check the correctness of a single constraint across the entire execution trace by evaluating committed polynomials at a single random point. However, a zkEVM circuit comprises millions of heterogeneous constraints over a trace of length $\traceLength$. Polynomial commitment schemes alone provide no protocol for reducing this entire constraint system to a small number of verifier queries with provable soundness. An Interactive Oracle Proof (IOP)~\cite{ben2016interactive, ben2018scalable} supplies this protocol layer, allowing the verifier to issue targeted challenges and queries across multiple committed polynomials in a structured, round-by-round manner. Formally, an IOP is an interactive protocol where in round $i$ the prover defines an oracle $\mathcal{O}_i$ that the verifier queries adaptively at chosen points. In zkEVM constructions, these oracles are realized via polynomial commitments: each $\mathcal{O}_i$ corresponds to a committed polynomial $f_i(x)$ with an evaluation interface returning pairs $(f_i(x_j), \zkproof_{x_j})$ for queries $x_j$, where $\zkproof_{x_j}$ certifies consistency with the commitment. The verifier then issues a challenge $c_i$, which the prover must incorporate into constructing the next oracle $\mathcal{O}_{i+1}$ (e.g., by using $c_i$ as a linear combination coefficient or evaluation point). This multiround interaction pins down the prover to correct polynomials while limiting the verifier to a small number of queries per round. The interactive structure amplifies soundness by forcing the prover to commit to each round's polynomial before seeing the next challenge, preventing adaptive responses to verifier queries. IOP-based proof systems reduce verification of entire constraint systems to logarithmic work, typically $O(\log \traceLength)$~\cite{ben2016interactive, thaler2022proofs} or $O(\log^2 \traceLength)$~\cite{ben2018scalable}, using polynomial commitments and the Schwartz--Zippel lemma as building blocks. 

The multiround interaction in IOPs limits deployment in decentralized environments like blockchains, where zkEVM systems require non-interactive proofs. The \textit{Fiat--Shamir heuristic}~\cite{fiat1986prove} eliminates this interaction by letting the prover derive each challenge deterministically. In the interactive protocol, the verifier would issue challenge $c_i$ after receiving commitment $\commitment{f_i}$. Under the Fiat--Shamir transformation, the prover instead computes $c_i$ by hashing all prior commitments $\{\commitment{f_1}, \dots, \commitment{f_i}\}$ as shown in Equation~\ref{zkevm:eq:fiatshamir}, where $H$ is a cryptographic hash function modeled as a random oracle~\cite{ben2018scalable}. The prover repeats this for every round, collapsing the multiround IOP into a single non-interactive argument. Any verifier can recompute the same challenges from the published commitments and verify the proof independently. Because the prover must fix each commitment before the hash determines the corresponding challenge, soundness is preserved under the random oracle model~\cite{thaler2022proofs}.

\begin{equation}
    c_i = H(\commitment{f_1} \| \commitment{f_2} \| \dots \| \commitment{f_i})
    \label{zkevm:eq:fiatshamir}
\end{equation}

With witness polynomials encoding the execution trace, the remaining question is how to structure the constraint polynomials applied to them. A \textit{constraint system} specifies this structure, defining the form each constraint polynomial takes and how it combines with witness polynomials to enforce correctness~\cite{thaler2022proofs}. The rest of this section examines dominant constraint systems, notably Rank-1 Constraint System (R1CS) \cite{thaler2022proofs}, PLONKish \cite{gabizon2019plonk}, and the Algebraic Intermediate Representation (AIR) \cite{ben2018scalable}.

\subsection{Rank-1 Constraint System (R1CS)}
\label{zkevm:subsection:r1cs}
Rank-1 Constraint Systems (R1CS) \cite{thaler2022proofs} represent computations as collections of quadratic constraints over a finite field $\field$, forming the foundation for early SNARKs~\cite{groth2016size}. Each R1CS constraint enforces that the product of two linear combinations equals a third, as formalized in Equation~\ref{zkevm:eq:r1cs}. The \emph{witness vector} $\witness = (w_1, \dots, w_n)$ contains the computational values satisfying the constraints, extended to $(1, w_1, \dots, w_n)$ with $w_0 = 1$ to enable constant terms in linear combinations. Coefficient vectors $\mathbf{a}, \mathbf{b}, \mathbf{c} \in \field^{n+1}$ define the constraint structure. The inner product notation $\langle \mathbf{a}, \witness \rangle$ computes the linear combination, i.e.,  $\langle \mathbf{a}, \witness \rangle = \sum_{i=0}^n a_i w_i$.

\begin{equation}
        \langle \mathbf{a}, \witness \rangle \cdot \langle \mathbf{b}, \witness \rangle = \langle \mathbf{c}, \witness \rangle \label{zkevm:eq:r1cs} 
\end{equation}

To illustrate, encoding $c = a + b$ in R1CS uses witness vector $\witness = (1, a, b, c)$ with coefficient vectors shown in Equation~\ref{zkevm:eq:r1cs:add1}. Applying Equation~\ref{zkevm:eq:r1cs} yields the constraint $(a + b) \cdot 1 = c$ as shown in Equation~\ref{zkevm:eq:r1cs:add2}. This minimal example omits the bit-width bounds, range checks, and overflow handling required for actual EVM operations. When a circuit must enforce multiple operations, R1CS introduces \emph{selector variables} $\selector{j} \in \{0,1\}$ into $\witness$, one per operation. Each constraint is multiplied by its selector, causing it to vanish algebraically when inactive. Consider the EVM \texttt{JUMPI} opcode, which conditionally sets the program counter to a target address or advances it sequentially. Wrapping its branch logic in a selector $\selector{\texttt{jumpi}}$ that activates only when \texttt{JUMPI} executes yields Equation~\ref{zkevm:eq:r1cs:jumpi}, where $\mathsf{cond} \in \{0,1\}$ is the branch condition and $\mathsf{pc'}$ is the next program counter value. The constraint holds trivially when $\selector{\texttt{jumpi}} = 0$, deactivating the opcode at steps where \texttt{JUMPI} does not execute.

\begin{align}
    &\mathbf{a} = (0, 1, 1, 0), \quad \mathbf{b} = (1, 0, 0, 0), \quad \mathbf{c} = (0, 0, 0, 1) \label{zkevm:eq:r1cs:add1}\\
    &\langle \mathbf{a}, \witness \rangle = a + b, \quad \langle \mathbf{b}, \witness \rangle = 1, \quad \langle \mathbf{c}, \witness \rangle = c \label{zkevm:eq:r1cs:add2} \\
    &\selector{\texttt{jumpi}} \cdot \big((1 - \mathsf{cond}) \cdot (\mathsf{pc'} - \mathsf{pc} - 1) + \mathsf{cond} \cdot (\mathsf{pc'} - \mathsf{target})\big) = 0
    \label{zkevm:eq:r1cs:jumpi}
\end{align}

R1CS proves inadequate for zkEVM implementation due to constraint explosion rooted in three structural limitations. The first is global constraint evaluation. An R1CS instance is a flat system of simultaneous equations over a single shared witness vector $\witness$. Every constraint is an equation over the full witness vector $\witness$: its linear combinations span all variables, not just those relevant to the operation being constrained. There is no native notion of execution step or conditional evaluation. 

The second limitation is the degree-2 ceiling of Equation~\ref{zkevm:eq:r1cs}: each constraint admits at most one multiplication between two linear combinations. Operations requiring higher-degree expressions must be split into multiple constraints with auxiliary intermediate variables. Equation~\ref{zkevm:eq:r1cs:jumpi} illustrates this limitation: it contains three multiplications (two bilinear products in the inner expression and one outer selector multiplication). Together they exceed the single-multiplication limit of one R1CS constraint, so encoding \texttt{JUMPI} requires three separate constraints with auxiliary variables. This degree ceiling amplifies across complex opcodes: a faithful 256-bit \texttt{ADD} requires 200--300 constraints for bit decomposition, range checks, and field arithmetic normalization~\cite{belles2022circom}, while \texttt{SHA3} requires approximately $150$K constraints for Boolean logic emulation and bitwise operations~\cite{belles2022circom}.

The third limitation is that non-arithmetic operations, those whose semantics involve bit-level or byte-level representations rather than native field arithmetic, must be emulated entirely through polynomial constraints. Range checks, opcode decoding, and byte decompositions have no dedicated mechanism in R1CS. To illustrate, decomposing a $256$-bit field element $x \in \field$ into $32$ bytes $\{x_{31}, \dots, x_0\}$ with $x = \sum_{i=0}^{31} x_i \cdot 256^{i}$ requires bit-decomposing each byte as $x_i = \sum_{j=0}^7 b_{i,j} 2^j$, with each $b_{i,j} \in \{0,1\}$. This demands $256$ Booleanity constraints $b_{i,j}(1 - b_{i,j}) = 0$, $32$ byte-recomposition constraints, and one top-level recomposition, totaling $289$ constraints for a single value. With 140+ opcodes potentially invoked repeatedly per transaction, naive R1CS circuits accumulate tens of millions of constraints per batch of transactions. These expressiveness limitations render R1CS impractical for production zkEVMs, motivating the adoption of more expressive frameworks like PLONKish and AIR arithmetizations.

\subsection{PLONKish Arithmetization}
\label{zkevm:subsec:plonkish}
PLONKish arithmetization~\cite{gabizon2019plonk} represents the witness as an execution trace table $\fulltrtab$, with $\traceLength$ rows for execution steps and $\witnessColumns$ columns for witness variables. Each row $\traceRow{i}$ captures the complete virtual machine state at step $i$. Each witness column tracks a specific witness variable across steps, e.g., a stack slot, a memory cell, the opcode, or program counter. The notation $\traceColRow{a}{i}$ denotes the value of column $a$ at row $i$. The circuit (constraint set) specifies polynomial constraints applied to this table: within-row constraints enforce gate semantics (e.g., $\traceColRow{a}{i}\cdot \traceColRow{b}{i}=\traceColRow{c}{i}$ for multiplication), and cross-row constraints capture state transitions (e.g., $\pcTrace{i+1}=\pcTrace{i}+1$ for program counter). This tabular organization directly addresses the global evaluation problem of R1CS (Section~\ref{zkevm:subsection:r1cs}), where every constraint spans the entire witness vector. The row-wise structure confines each constraint to individual computation steps, operating on $\witnessColumns$ witness variables per row rather than the full $\traceLength \times \witnessColumns$ witness.

PLONKish arithmetization enforces constraints row-wise (not globally across all rows) using \emph{selector polynomials}, which activate specific gate logic conditionally in each row. Conditional constraint activation in PLONKish follows a similar principle to R1CS. Selector values indicate which opcode executes at each row, activating only the corresponding constraint. The efficiency advantage over R1CS comes from the tabular trace structure, which confines constraint evaluation to individual rows rather than checking every constraint against the entire witness. For example, consider enforcing the semantics of the \texttt{ADD} opcode. Each row $i$ of the trace corresponds to a single EVM execution step and contains columns such as the current opcode $\opcodeTrace{i}$, the top two stack values $\traceColRow{a}{i}, \traceColRow{b}{i}$, and the stack result $\traceColRow{c}{i}$. This is naively constrained by a PLONKish selector, which states that when the opcode equals \texttt{ADD}, the constraint $\traceColRow{a}{i} + \traceColRow{b}{i} = \traceColRow{c}{i}$ must hold. This is encoded using a selector polynomial $\selectorAt{\texttt{add}}{i} \in \{0,1\}$, which activates the constraint only when the opcode matches, i.e., $\opcodeTrace{i} = \texttt{ADD}$ as shown by Equation~\ref{zkevm:eq:plonkish:add}. The constraint is enforced only on the rows where $\selectorAt{\texttt{add}}{i} = 1$, and remains trivially satisfied otherwise. 

\begin{equation}
    \selectorAt{\texttt{add}}{i} \cdot \big(\traceColRow{a}{i} + \traceColRow{b}{i} - \traceColRow{c}{i}\big) = 0 \label{zkevm:eq:plonkish:add}
\end{equation}

\emph{Custom gates} in PLONKish resolve the degree-2 ceiling of R1CS (Section~\ref{zkevm:subsection:r1cs}) by allowing the circuit designer to define polynomial constraints of degree greater than two~\cite{chen2022hyperplonk}. A single higher-degree constraint can capture logic that would otherwise require decomposition into multiple constraints with auxiliary intermediate variables. As an example, consider the EVM \texttt{JUMPI} opcode, which conditionally updates the program counter. A custom gate enforces that if $\traceColRow{cond}{i} \in \{0,1\}$ is true, the program counter jumps to $\traceColRow{target}{i}$; otherwise, it advances sequentially. As shown by Equation~\ref{zkevm:eq:plonkish:jumpi}, this behavior is encoded in a single constraint activated by a selector polynomial $\selectorAt{\texttt{jumpi}}{i}$, where $\pcTrace{i}$ denotes the program counter at step $i$. Equation~\ref{zkevm:eq:plonkish:jumpi} is a degree-3 expression. Recall from Equation~\ref{zkevm:eq:r1cs:jumpi} that encoding this same logic in R1CS requires three separate constraints with auxiliary intermediate variables, since each R1CS constraint admits at most one bilinear product. The custom gate collapses all three into a single constraint, directly resolving the degree-2 ceiling of R1CS.

\begin{equation}
    \selectorAt{\texttt{jumpi}}{i} \cdot \big((1 - \traceColRow{cond}{i}) \cdot (\pcTrace{i+1} - \pcTrace{i} - 1) + \traceColRow{cond}{i} \cdot (\pcTrace{i+1} - \traceColRow{target}{i})\big) = 0
    \label{zkevm:eq:plonkish:jumpi}
\end{equation}

\emph{Lookup arguments} enable PLONKish systems to efficiently handle the non-arithmetic operations that require costly bit-level emulation in R1CS (Section~\ref{zkevm:subsection:r1cs})~\cite{gabizon2020plookup}. A lookup argument proves that each witness value $w_i$ in a set $W = \{w_1, \dots, w_n\}$ appears in a predefined table $T = \{t_1, \dots, t_m\}$. The proof reduces this set-membership claim to a polynomial equality check, verified non-interactively via the Fiat--Shamir heuristic~\cite{fiat1986prove} with negligible soundness error by the Schwartz--Zippel lemma~\cite{schwartz1980fast}. Returning to the byte decomposition example from Section~\ref{zkevm:subsection:r1cs}, which requires $289$ R1CS constraints, a lookup argument proves each of the $32$ bytes lies in table $T = \{0, \dots, 255\}$. This requires $32$ lookup constraints plus one recomposition constraint, totaling just $33$ constraints, nearly $9$ times fewer. This reduction is in constraint count. Each lookup is costlier to prove than a single R1CS check. However, the lookup protocol pays a one-time setup cost shared across all lookups, making each additional lookup inexpensive~\cite{gabizon2020plookup}. Critically, lookup arguments verify inclusion but not semantic validity. Confirming that every $w_i$ appears in $T$ does not guarantee that $T$ encodes correct relationships. For the byte range table $T = \{0, \dots, 255\}$, correctness is trivial since the table is a fixed public constant. For more complex operations, correctness requires proof. Consider \texttt{SHA3} (Keccak-256) hashing, which requires approximately $150$K constraints in R1CS~\cite{belles2022circom}. Rather than embedding this cost at every call site, an auxiliary circuit can prove a table of $(x, \mathsf{hash}(x))$ pairs once during proof generation. The main zkEVM circuit then verifies hash operations through lookups into this validated table, paying only one lookup constraint per invocation.

\emph{Permutation arguments} in PLONKish systems enforce copy constraints across the trace~\cite{gabizon2019plonk}. A copy constraint requires that two cells at different positions hold the same value. For example, suppose \texttt{ADD} at row $i$ produces a result $\traceColRow{c}{i}$ that \texttt{MUL} at row $j$ needs as an operand $\traceColRow{a}{j}$. The constraint $\traceColRow{c}{i} = \traceColRow{a}{j}$ must then hold. In zkEVMs, such constraints enforce dataflow equivalence. Stack outputs wire to opcode inputs, and memory writes match subsequent reads. Enforcing each such pair as a separate polynomial constraint would add thousands of constraints to the circuit. The permutation argument avoids this by defining a single rearrangement that swaps all paired cells simultaneously. In the example above, the permutation argument swaps position $(i, c)$ with position $(j, a)$, and similarly for every other pair that must be equal. The argument then checks whether the entire trace remains unchanged under this rearrangement. If all paired cells hold equal values, swapping them has no effect and the check passes. If any pair differs, the swap alters the trace and the check fails. This replaces thousands of individual equality constraints with a single polynomial equality check~\cite{gabizon2019plonk}. PLONKish is the dominant arithmetization in production zkEVMs, owing to the expressiveness of custom gates, lookup arguments, and permutation arguments~\cite{scroll2024whitepaper, zksync2024protocol, matterlabs2024boojum, linea2022, taiko2024circuits}. AIR provides an alternative arithmetization based on uniform state-transition constraints, forming the basis for STARK-based proof systems.

\subsection{Algebraic Intermediate Representations (AIR)}
\label{zkevm:subsec:air}
AIR defines computation through a fixed set of transition constraints $\{C_1, \ldots, C_s\}$ applied uniformly to every consecutive row pair in the trace table $\fulltrtab$~\cite{ben2018scalable}. Each constraint enforces $C_j(\traceRow{i}, \traceRow{i+1}) = 0$, referencing columns within row $i$, row $i+1$, or both. This captures both within-row operation semantics and cross-row state transitions in a single expression. Constraint activation in AIR relies on binary flag columns that function identically to the selector polynomials of PLONKish~\cite{goldberg2021cairo,miden_whitepaper}. The structural distinction of the AIR formalism from PLONKish lies in two absent primitives, namely native lookup arguments and native permutation arguments. Without these, the formalism must encode set membership and copy constraints algebraically within transition constraints, sharing the third limitation of R1CS (Section~\ref{zkevm:subsection:r1cs}).

The AIR formalism compensates for the absence of native lookups through the cost model of the FRI commitment scheme~\cite{ben2018scalable}. To verify $x \in \{0, \dots, 255\}$, PLONKish invokes a single lookup into a precomputed table~\cite{gabizon2020plookup}. The formalism instead defines a vanishing polynomial $P(x)$ per Equation~\ref{zkevm:eq:air:range}. This polynomial equals zero if and only if $x$ lies in the target set. The resulting degree is 256. Under KZG commitments~\cite{kate2010constant}, this is prohibitive as committing to a degree-$d$ polynomial requires $d+1$ group exponentiations. FRI-based commitment, however, scales quasi-linearly using only hash operations over Merkle trees. Since hashes are far cheaper than exponentiations, FRI makes high-degree vanishing polynomials practical. The formalism thus offsets the absence of native lookups at the cost of higher-degree constraints.

\begin{equation}
P(x) = \prod_{i = 0}^{255} (x - i)
\label{zkevm:eq:air:range}
\end{equation}

In practice, PLONKish arithmetization underpins most production zkEVMs~\cite{scroll2024whitepaper, zksync2024protocol, matterlabs2024boojum, linea2022, taiko2024circuits}. The native lookup and permutation arguments of PLONKish align naturally with the diverse requirements of the 140+ EVM opcodes. Range checks, byte decompositions, memory consistency, and stack wiring each benefit from dedicated argument types. Polygon zkEVM instead builds on AIR, adding inclusion and permutation arguments to the constraint system~\cite{polygon_zkevm}. Purpose-built zkVMs pair AIR with FRI-based STARKs~\cite{goldberg2021cairo, miden_whitepaper}. These systems design compact instruction sets, a handful of canonical instruction forms encoded through 15 one-bit flags in Cairo~\cite{goldberg2021cairo} and roughly 81 basic-block operations in Miden~\cite{miden_whitepaper}. Their opcodes share similar algebraic structure, which a uniform transition constraint captures directly.

Table~\ref{zkevm:tab:arithmetization-comparison} summarizes the core differences between R1CS, PLONKish, and AIR arithmetizations. R1CS laid foundations but lacks expressiveness for production zkEVMs \cite{belles2022circom, groth2016size}. PLONKish dominates deployed implementations~\cite{scroll2024whitepaper, zksync2024protocol, matterlabs2024boojum, linea2022, taiko2024whitepaper, taiko2024circuits}. AIR pairs naturally with compact instruction sets of zkVMs~\cite{goldberg2021cairo,miden_whitepaper}.

\begin{table}[t]
\centering
\renewcommand{\arraystretch}{1.3}
\footnotesize 
\begin{tabularx}{\textwidth}{|>{\columncolor{gray!15}\bfseries}l|>{\raggedright\arraybackslash}p{2.1cm}|X|X|}
\rowcolor{gray!25}
\hline
\textbf{Feature} & \textbf{R1CS} \cite{thaler2022proofs} & \textbf{PLONKish} \cite{gabizon2019plonk} & \textbf{AIR} \cite{ben2018scalable} \\
\hline
Constraint form      & Bilinear products & Polynomial identities     & Transition constraints \\
\hline
Constraint scope     & Global (full witness)     & Selector-gated rows & Consecutive row pairs \\
\hline
Set membership      & Bit decomposition & Native lookups      & Vanishing polynomials \\
\hline
Copy constraints & Algebraic encoding           & Native (permutation arguments) & Algebraic encoding \\
\hline
Constraint degree cost & Fixed at 2 (bilinear)     & Linear in exponentiations       & Quasi-linear in hashing \\
\hline
Constraint specialization & Uniform (single bilinear form) & Per-operation (custom gates) & Uniform (flag-differentiated) \\
\hline
Base formalism          & \xmark   & \cmark (Scroll~\cite{scroll2024whitepaper}, zkSync~\cite{zksync2024protocol, matterlabs2024boojum}, Linea~\cite{linea2022}, Taiko~\cite{taiko2024whitepaper, taiko2024circuits})     & \cmark (Polygon zkEVM~\cite{polygon_zkevm}, and the zkVMs Cairo~\cite{goldberg2021cairo}, Miden~\cite{miden_whitepaper}) \\
\hline
\end{tabularx}
\caption{Comparison of R1CS, PLONKish, and AIR in zkEVM design. The constraint specialization row denotes whether each operation type uses a dedicated constraint or shares a uniform template differentiated by flags.}
\Description{A comparison table of three arithmetization approaches (R1CS, PLONKish, and AIR) across seven features: constraint form, constraint scope, set membership encoding, copy constraint mechanisms, constraint degree cost, constraint specialization, and base formalism. The table shows that PLONKish is the base formalism of four production zkEVMs (Scroll, zkSync, Linea, Taiko), that AIR is the base formalism of Polygon zkEVM and of the zkVMs Cairo and Miden, and that R1CS is the base formalism of none of them.}
\label{zkevm:tab:arithmetization-comparison}
\end{table}

\section{Opcode Dispatch Mechanisms: From Static Constraint Inlining to Dynamic Execution}
\label{zkevm:sec:evolution}
Static constraint inlining is the most direct approach to encoding an EVM program as a ZKP circuit. We refer to an EVM \emph{program} as the bytecode of a smart contract when executed at EVM runtime. A single transaction may execute one or more programs depending on which contracts it invokes, and a zkEVM proves the correct execution of these programs. In the static inlining approach, the prover builds a specialized circuit for each specific execution trace, one that hardcodes exactly the opcodes that execute, in the order they execute. If a program runs \texttt{ADD}, then \texttt{MUL}, then \texttt{SHA3}, the prover constructs a circuit that inlines these three opcodes in sequence. Every execution thus requires building a new circuit from scratch, repeating all circuit-dependent preprocessing, and communicating the circuit structure to the verifier before verification can begin.

Such static constraint inlining fails three universal requirements that any deployed zkEVM must satisfy. The first requirement is prover scalability. The prover preprocesses the circuit into an internal representation it reuses across many proofs. A per-execution circuit forces this preprocessing to be redone for every proof, which defeats caching and prevents a prover fleet from pipelining proofs of distinct executions in parallel. Both capabilities are necessary to meet production proof throughput. The second requirement is on-chain verification. A zkEVM rollup deploys a single verifier contract on Ethereum bound to one fixed circuit, and this contract cannot be updated per transaction or per block. Static inlining leaves no stable verifier to deploy. The third requirement is trust in the circuit itself. A zkEVM proof demonstrates only that some execution satisfies the circuit constraints. Its semantic meaning as a valid EVM execution depends on the circuit faithfully encoding the program within EVM semantics. That trust is credible only when a single canonical circuit is publicly audited and endorsed by the ecosystem, so that every verifier relies on the same openly agreed-upon artifact. A per-execution circuit cannot serve as this shared artifact. Each execution produces a different circuit, so there is no fixed object for the ecosystem to audit once and rely on across all subsequent proofs. SNARK-based deployments impose a fourth requirement. The trusted setup that SNARKs require is prohibitively expensive to repeat, so it must be performed once and reused across all future proofs.

Meeting these requirements forces a single, predetermined circuit that is reused across all executions. A canonical circuit cannot hardcode any particular program because it must be capable of accommodating every possible opcode at every step of the execution trace. This is done by replicating the constraint set of each opcode at every execution step (i.e., a row in PLONKish and AIR arithmetizations), even though exactly one opcode is active per step. The prover must then enforce the constraints of the active opcode at each step and suppress all others. The mechanism that performs this step-by-step activation is \emph{opcode dispatch}, which is implemented using selectors.

\subsection{Selector-Based Opcode Dispatch}
\label{zkevm:subsec:selector-dispatch}

Selectors are the constraint-activation machinery of a universal zkEVM circuit~\cite{taiko2024circuits}. The circuit predefines all possible constraints during compilation and replicates them at every execution step; selectors then dynamically choose which constraints to activate at each step~\cite{gabizon2019plonk, scroll2024whitepaper}. The execution trace table $\trtab$ captures EVM state transitions, with each row corresponding to an opcode execution and columns encoding state components.
Each opcode is modeled by a low-degree constraint polynomial $C_j$ over $\field$ whose variables reference trace table $\trtab$ entries. $C_j(i) = 0$ when constraint $j$ is satisfied at row $i$. 

The dispatch mechanism uses selector columns in the trace table, where each selector $\selector{j}$ forms a dedicated column $\traceCol{s_j}$ with binary values~\cite{gabizon2019plonk}. The witness polynomial $\selectorPoly{j}$ is interpolated from this column $\traceCol{s_j}$, equaling 1 at row $i$ where opcode $j$ executes ($\traceColRow{s_j}{i} = 1$) and 0 elsewhere ($\traceColRow{s_j}{i} = 0$). At each row $i$, the one-hot constraint $\sum_{j=1}^{\opcodeCount} \traceColRow{s_j}{i} = 1$ ensures exactly one selector is active, activating the corresponding constraint $C_j$ for that execution step~\cite{gabizon2019plonk}. With $\opcodeCount$ supported opcodes, the dispatch enforces the constraint $\sum_{j=1}^\opcodeCount \traceColRow{s_j}{i} \cdot C_j(i) = 0$ at each trace row $i$, where only the active gate contributes since only one $\traceColRow{s_j}{i} = 1$. This transforms constraint activation from compile-time to runtime dispatch~\cite{gabizon2019plonk}. 

Deployed zkEVM circuits reduce selector overhead by grouping opcodes with shared structure under a single selector~\cite{taiko2024circuits}. Rather than one selector per opcode, related opcodes are assigned to $\groupNum$ groups with $\groupNum < \opcodeCount$, each carrying a group selector $\selector{g}$ and a shared constraint polynomial $C_g$ that captures the common structure of its members. Opcodes such as \texttt{ADD} and \texttt{SUB}, for instance, can be covered by one group selector whose constraint encodes both operations. This preserves the dispatch constraint $\sum_{g=1}^{\groupNum} \traceColRow{s_g}{i} \cdot C_g(i) = 0$ while reducing both trace width and the number of selector polynomial commitments from $\opcodeCount$ to $\groupNum$. The approach is enabled by the custom gate capability of PLONKish, which admits higher-degree constraints that express compound operations directly rather than decomposing them into many basic gates~\cite{gabizon2019plonk, scroll2024whitepaper}.

\subsection{Program-Binding via Bytecode Tables and ROM}
\label{zkevm:subsec:rom-based}

A universal zkEVM circuit that runs arbitrary programs must bind the execution trace to the specific program being executed. The selector mechanism binds the trace to a series of opcodes, and each constraint polynomial $C_j(i) = 0$ enforces that whichever opcode fires at row $i$ executes correctly. But nothing in this mechanism ties the series of opcodes to the EVM bytecode that the transaction actually invokes. Without such a binding, a satisfying proof attests only that some sequence of opcodes is internally consistent, not that the sequence matches the invoked program. A prover could truthfully prove a \texttt{SUB} at row $i$ with correctly updated stack values, while the program specified \texttt{ADD} at that step. The constraint polynomial for \texttt{SUB} is satisfied, yet the trace has diverged from the actual program that EVM executed. Closing this gap outside the circuit would require the verifier to inspect the execution trace row by row and match each row against the bytecode of the program, which defeats the point of succinct verification. zkEVMs through Type 2 instead fold this check into the circuit by encoding the program as a public Read-Only Memory ($\rom$), a table $\{(j, r_j)\}_{j=0}^{\progLen-1}$ with $\progLen$ rows, mapping each program-counter address $j$ to a row $r_j$ that encodes instruction data. The execution trace is extended with a dedicated program-counter column alongside the selector columns, where $\pcTrace{i}$ records the program-counter value at row $i$. A lookup argument binds $(\pcTrace{i},\, r_{\pcTrace{i}}) \in \rom$ at each trace row, certifying the match as part of the proof. This is done \emph{in-circuit}, enforced by constraints rather than checked externally by the verifier~\cite{gabizon2020plookup}. Because $\pcTrace{i}$ is a witness column, control flow is dynamic: only executed rows appear in the trace, and untaken branches never materialize. The $\rom$ itself is part of the public statement that must correspond to the on-chain deployed bytecode, ensuring the $\rom$ table reflects the correct program. Together, $\pcTrace{i}$ determines \emph{which} instruction is fetched from $\rom$, while the selector columns determine \emph{how} the fetched instruction is activated as constraints.

zkEVMs through Type 2 use two $\rom$ designs that differ in what each row $r_j$ stores. In \emph{explicit} $\rom$ systems such as Polygon zkEVM~\cite{polygon_zkevm} and Linea~\cite{linea2022,begassat2021specification}, each $r_j$ holds the selector values and operand-routing flags of the opcode at address $j$, precomputed from the raw bytecode at $\rom$-build time rather than in-circuit. At trace row $i$ with $\pcTrace{i} = j$, a lookup $(j, r_j) \in \rom$ binds the row to the program. In \emph{implicit} $\rom$ systems such as Scroll~\cite{scroll2024whitepaper} and Taiko~\cite{taiko2024whitepaper}, each $r_j$ stores the raw bytecode byte at address $j$, and the circuit derives selector values and opcode groups from that byte in-circuit, as part of the proof generation. The explicit $\rom$ table is therefore wide, each row a vector of decoded selectors, but adds no prover overhead for decoding the raw bytecode in-circuit. The implicit $\rom$ table is narrow, each row just the raw bytecode byte, but pushes the per-byte decode onto the prover, performed in-circuit. 

The two designs also differ in whether the decode is proven inside the circuit or assumed and audited outside it. In both, the public statement commits to the bytecode deployed on-chain. In the implicit design, $\rom$ holds that same deployed raw bytecode, so the hash of $\rom$ equals the bytecode hash the statement commits to. The $\rom$ lookup binds the trace to the program that the statement commits to, and the decode is enforced in-circuit, inside the proven relation. In the explicit design, the lookups instead consult a $\rom$ that holds the decoded form of the program, not the raw bytecode deployed on-chain. Its hash therefore differs from the bytecode hash the statement commits to, since the hash is taken over the decoded selectors, and not over the raw bytecode. Decoding the raw bytecode into selectors is performed off-circuit, by a build step or interpreter that must itself be trusted or independently audited. Hence, the proof certifies execution of the decoded $\rom$, not that this $\rom$ faithfully decodes the deployed bytecode. The implicit design thus proves the decode within the circuit, while the explicit design assumes it and delegates trust to the decoder.

Table~\ref{tab:zkevm-rom-activation-hybrid} summarizes the three approaches. The evolution from static inlining to universal circuits with program-table lookups is the architectural step that distinguishes production zkEVMs from program-specific circuits. The implicit-versus-explicit choice is a refinement within that regime. Designs such as zkSync Era \cite{zksync2024protocol} lie outside this spectrum as they execute a non-EVM instruction set rather than the EVM opcodes.

\begin{table}[t]
\centering
\renewcommand{\arraystretch}{1.3}
\footnotesize
\begin{tabularx}{\textwidth}{|>{\columncolor{gray!15}\bfseries}l|
>{\centering\arraybackslash}X|
>{\centering\arraybackslash}X|
>{\centering\arraybackslash}X|}
\hline
\rowcolor{gray!25}
Attribute & Static Inlining & Implicit ROM & Explicit ROM \\
\hline
ROM Content & None (program embedded in circuit) & Raw bytecode bytes & Selector values and flags \\
\hline
Dispatch Decoding & None (hardcoded) & In-circuit (at proof time) & In ROM (at ROM-build time) \\
\hline
Program-Counter Role & Implicit in circuit structure & Witness (PC-indexed lookup) & Witness (PC-indexed lookup) \\
\hline
Constraint Activation & Hardcoded (no selectors) & Selectors & Selectors \\
\hline
Control Flow & Static (fixed path) & Dynamic PC-driven & Dynamic PC-driven \\
\hline
Branch Handling & All paths inlined & Only taken path & Only taken path \\
\hline
Program Auditability & Circuit encodes one program & Bytecode visible & Pre-decoded dispatch in $\rom$ \\
\hline
zkEVM Projects & \xmark & Scroll~\cite{scroll2024whitepaper}, Taiko~\cite{taiko2024whitepaper} & Polygon~\cite{polygon_zkevm}, Linea~\cite{linea2022,begassat2021specification} \\
\hline
\end{tabularx}
\caption{Comparison of program-binding approaches across the zkEVM design space. Static inlining embeds the program in the circuit; zkEVMs through Type 2 use a universal circuit with a program-counter-indexed lookup against a public ROM, differing in whether dispatch decoding is pre-populated in the table (explicit ROM) or derived in the circuit (implicit ROM).}
\Description{A comparison table of three program-binding approaches (static inlining, implicit ROM, and explicit ROM) across eight attributes: ROM content, dispatch decoding location, program-counter role, constraint activation, control flow, branch handling, program auditability, and example projects. Static inlining has no ROM and encodes one program per circuit; both production strategies use dynamic PC-driven control flow and selectors for constraint activation, differing in where dispatch decoding lives.}
\label{tab:zkevm-rom-activation-hybrid}
\end{table}

\section{Zero-knowledge friendly rewrites of EVM semantics}
\label{zkevm:sec:rewrites}

zkEVM circuits encode EVM semantics as low-degree polynomial constraints over an execution trace, but the canonical execution model of EVM does not naturally fit such a trace. The operand stack of EVM is a last-in-first-out (LIFO) buffer of up to $1024$ live values~\cite{wood2014ethereum}. A direct trace encoding dedicates one column per stack slot, inflating trace width by up to $1024$ columns. Each column must be interpolated and committed separately, raising the witness polynomial count from $\witnessColumns$ to $\witnessColumns + 1024$ even though typical programs touch only a small portion of that space. Selectors can disable unused stack-column constraints, but each column remains committed as part of the witness. 

Additionally, memory and persistent storage require every read to return the value most recently written at that address, an invariant known as \emph{read-after-write consistency}. PLONKish constraints are \emph{row-local} in the sense that they can reference rows at any fixed offset chosen at circuit compile time. However, for the read-after-write consistency, the matching write may sit any number of rows earlier, with the exact distance determined only at runtime. Covering all possible distances naively would require a separate constraint polynomial for every offset $1, 2, \ldots, \traceLength-1$ at every row. Each is activated by a selector only at the few rows that read from that offset, while most rows activate none. Yet PLONKish evaluates every constraint polynomial at every row, inflating circuit size $\circuitSize$ from $\Theta(\traceLength)$ to $\Theta(\traceLength^2)$ for read-after-write enforcement alone~\cite{liu2025ceno}.

Complex opcodes such as \texttt{SHA3}, \texttt{SLOAD}, \texttt{SSTORE}, \texttt{CALL}, and some of the precompiles mix bit-level, hash, modular, and elliptic-curve operations that no native field constraint expresses compactly. When an opcode does not fit native field constraints, the canonical encoding must decompose every $256$-bit EVM word into many smaller columns, such as bit columns for \texttt{SHA3}. A single \texttt{SHA3} or precompile invocation can then generate enough such constraints to dominate the main trace.

Encoding canonical EVM execution directly therefore inflates trace width, circuit size, and per-row constraint count regardless of arithmetization choice. To overcome these inefficiencies, this section examines how semantic rewrites transform these patterns into zero-knowledge-friendly forms. The transformations reduce stack column overhead, lift memory consistency into specialized subcircuits, and decompose complex opcodes into low-degree sub-operations.
\subsection{Stack and Memory Modeling in zkEVMs}
\label{zkevm:sec:rewrites:subsec:stack-and-memory}
Specialized stack and memory modules address the overheads noted earlier by lifting stack and memory state into auxiliary subcircuits dedicated to enforcing their invariants outside the main execution trace. The stack module commits two tables as part of the witness. The first is the stack table $\stkTab$ whose rows $(\mathsf{sp}, \mathsf{val}, \mathsf{op}, \mathsf{timestamp})$ log every stack access of the main execution trace. Here $\mathsf{sp}$ is the stack-pointer position and $\mathsf{val}$ is the value read or written. The flag $\mathsf{op}$ distinguishes reads from writes. The field $\mathsf{timestamp}$ is bound to a counter column that the prover commits in the main trace. A row-local constraint in the main trace forces this counter to advance by exactly one at every row that emits a stack access. The counter remains unchanged at rows without a stack access, and therefore strictly increases across access-emitting rows. At main-trace step $i$ it has a unique value $c_i$ determined by the trace up to that point. When step $i$ of the main trace emits a stack access, the prover appends a row to $\stkTab$ whose timestamp slot holds $c_i$. This binds the $\stkTab$ row to main-trace step $i$. The second table, $\stkTab_{\texttt{sorted}}$, is a sorted version of $\stkTab$ holding the same rows reordered by $(\mathsf{sp}, \mathsf{timestamp})$, so that consecutive rows at the same stack-pointer position sit side by side. Three proof obligations bind these tables together. A permutation argument shows that the set of stack accesses emitted by the main trace equals the set of rows in $\stkTab$. This pins both the contents and the cardinality of $\stkTab$ to the main trace, preventing the prover from padding $\stkTab$ with rows that do not correspond to any main-trace access. A second permutation argument shows that $\stkTab_{\texttt{sorted}}$ is a reordering of $\stkTab$. Row-local constraints inside the stack module then enforce the sort of $\stkTab_{\texttt{sorted}}$ and the read-after-write relation.

Specifically, the row-local relation at $\stkTab_{\texttt{sorted}}$ requires that for every read row $(\mathsf{sp}_i, \mathsf{val}_i, \mathsf{read})$, the value $\mathsf{val}_i$ equals the value of the most recent prior write at the same stack-pointer position. In $\stkTab$ this prior write may sit any number of rows earlier, exactly the inherent row-local mismatch of PLONKish noted earlier. Sorting by $(\mathsf{sp}, \mathsf{timestamp})$ collapses the gap. In $\stkTab_{\texttt{sorted}}$, all accesses to a single stack-pointer position form a contiguous block ordered by timestamp. Within each block, the immediately preceding row is, by virtue of the sort, the closest prior access at that position. The row-local check therefore enforces that every read row carries the same value as this immediate predecessor, which is either the most recent read or the most recent write. By induction along the block, every read returns the value most recently written at that position, even when many reads sit between it and the write. This is the read-after-write invariant enforced by the stack module. Such an adjacent-row check is precisely what PLONKish constraints handle natively. Because real programs touch only a small portion of the stack, $\stkTab$ commits one row per actual access rather than $1024$ columns per main-trace row, eliminating the column-count overhead~\cite{scroll2024whitepaper, polygon_zkevm}. LIFO ordering and the full $1024$-element stack capacity of EVM are preserved unchanged. Stack-depth bounds are enforced by per-opcode stack-pointer constraints in the main trace, separate from the value-consistency role of the stack module.

The memory module uses the same construction with $\memTab$ and $\memTab_{\texttt{sorted}}$ replacing $\stkTab$ and $\stkTab_{\texttt{sorted}}$, and $\mathsf{addr}$ (memory address) replacing $\mathsf{sp}$ (stack-pointer position). Each row therefore carries the tuple $(\mathsf{addr}, \mathsf{val}, \mathsf{op}, \mathsf{timestamp})$. The same row-local relation at $\memTab_{\texttt{sorted}}$, with blocks now grouped by $\mathsf{addr}$ instead of $\mathsf{sp}$, enforces the read-after-write invariant for memory. Every memory read returns the value most recently written at that address, by the same induction along the address-grouped block as in the stack case. Unwritten addresses default to zero per EVM semantics, encoded by implicit zero-initialization of $\memTab$ for first-touch reads. 

Because this module-based rewrite is confined to how stack and memory state are witnessed inside the proof system, the deployed bytecode and per-opcode runtime semantics are unchanged. Bytecode and opcode fidelity are therefore preserved, and the rewrite does not change the type classification of the zkEVM on its own. Scroll~\cite{scroll2024whitepaper, scroll2024circuits}, Polygon zkEVM~\cite{polygon_zkevm}, Linea~\cite{linea2022, begassat2021specification}, and Taiko~\cite{taiko2024circuits} all adopt this module-based design for stack and memory.

\subsection{Opcode Decomposition and Auxiliary Tables}
\label{zkevm:subsec:opcode-decomposition}

The most acute case among the rewrites motivated earlier in this section is the family of complex opcodes (\texttt{SHA3}, \texttt{CALL}, \texttt{SSTORE}, and several precompiles) that no single monolithic custom gate can efficiently absorb. A monolithic \texttt{SHA3} row would have to carry the $1600$-bit Keccak internal state ($\sim 25$ field elements), every intermediate value across the $24$ rounds of the Keccak-$f$ permutation~\cite{bertoni2009keccak}, and the associated Booleanity columns. These \texttt{SHA3}-specific columns would persist across every row of the main trace, including those where \texttt{SHA3} is inactive (Figure~\ref{zkevm:fig:sha3-opcode-decompose}.a). The same column persistence affects \texttt{CALL} and \texttt{SSTORE}, each of which touches multiple subsystems and would require its own dedicated witness layout. 

The opcode decomposition rewrite addresses this by splitting each complex opcode into low-degree sub-operations distributed across consecutive rows (Figure~\ref{zkevm:fig:sha3-opcode-decompose}.b). \texttt{SHA3} decomposes into per-round constraints over the $24$ rounds of Keccak-$f$~\cite{begassat2021specification, polygon_zkevm}, while \texttt{CALL} and \texttt{SSTORE} split into algebraically simple operations~\cite{scroll2024whitepaper, begassat2021specification}. The monolithic high-degree constraints hence become a family of low-degree sub-constraints with localized witness variables, and transition constraints link the sub-operations to ensure sequential integrity~\cite{polygon_zkevm, begassat2021specification}. 

Decomposition alone, however, does not yet minimize trace width because the intermediate columns still span every row of the main trace. For this reason, \emph{auxiliary tables} combine with opcode decomposition to achieve narrow main traces (Figure~\ref{zkevm:fig:sha3-opcode-decompose}.c). The main trace retains only input/output values while auxiliary tables handle complex logic. Each auxiliary table is specialized to one primitive (for example, one table for the Keccak round function, another for memory consistency, another for range checks), with constraints that enforce correctness over the rows of that table. Lookup arguments then bind main-trace values to corresponding auxiliary-table entries. For \texttt{SHA3}, they enforce that input/output values match the initial and final auxiliary states. Because these auxiliary constraints no longer have to be replicated and selector-zeroed at every row of the main trace, the per-row constraint count on the main trace shrinks accordingly. In addition, the auxiliary tables are interpolated over their own evaluation domain. Their size is determined by the volume of auxiliary work (for example, the number of Keccak invocations in the batch), not by the main-trace length. The witness polynomials over the auxiliary columns therefore have lower degree than the main-trace witness polynomials, lowering the witness polynomial commitment and interpolation cost on the auxiliary side. Hence, auxiliary tables contain only rows for actual opcode executions, keeping the main trace narrow.

A parallel family of rewrites targets the original precompiled contracts at fixed addresses \texttt{0x01} through \texttt{0x09}, where computationally expensive primitives such as modular exponentiation, BN254 elliptic-curve addition and scalar multiplication, and pairing checks are dispatched via \texttt{CALL}~\cite{wood2014ethereum}. Encoding these primitives as native field arithmetic over $\field$ inflates constraint counts because every 256-bit modular operand must be decomposed into smaller limbs that are range-checked individually, and each elliptic-curve formula expands into long sequences of multiplications. Linea therefore proves precompile execution separately from the main EVM trace~\cite{linea2022}.

The opcode decomposition and auxiliary table rewrites are confined to how each complex opcode is witnessed inside the proof system. Hence, the deployed bytecode and per-opcode input-output behavior are unchanged. Bytecode and opcode fidelity are therefore preserved, and neither rewrite changes the type classification of the zkEVM on its own. Production zkEVMs such as Linea~\cite{linea2022, begassat2021specification}, Polygon zkEVM~\cite{polygon_zkevm}, Scroll~\cite{scroll2024whitepaper}, and Taiko~\cite{taiko2024whitepaper} implement these rewrites. Type 2 zkEVMs additionally substitute proof-efficient internal algorithms for precompiles. Such substitutions preserve input-output behavior but alter both step count and gas cost~\cite{buterin_2022_zkevm}. This specific algorithmic change, not the auxiliary subcircuit routing, is what places those systems in Type 2 rather than Type 1.

\begin{figure*}[t]
\centering
{%
\ifpreprint
  \renewcommand{\scriptsize}{\fontsize{7}{8.4}\selectfont}%
  \renewcommand{\footnotesize}{\fontsize{8}{9.6}\selectfont}%
  \renewcommand{\small}{\fontsize{9}{10.8}\selectfont}%
  \renewcommand{\normalsize}{\fontsize{10}{12}\selectfont}%
\fi
\begin{tikzpicture}[scale=0.73, transform shape,
    cell/.style={
        rectangle,
        draw=black,
        minimum width=1.3cm,
        minimum height=0.6cm,
        font=\footnotesize,
        align=center
    },
    active/.style={
        cell,
        fill=gray!20
    },
    constraint/.style={
        trapezium,
        trapezium angle=80,
        draw=black,
        minimum width=3cm,
        minimum height=0.6cm,
        font=\small,
        align=center
    },
    trace_label/.style={
        font=\small,
        rotate=90
    },
    caption_style/.style={
        font=\normalsize,
        align=center
    }
]

\begin{scope}[shift={(-2.6,0)}]
    \node[constraint] at (1.45, 3) {High Degree SHA3 Constraint};

    \node[cell, font=\scriptsize] at (0, 0) {SHA3-1};
    \node[cell, font=\scriptsize] at (1.3, 0) {SHA3-2};
    \node[cell, font=\scriptsize] at (2.6, 0) {...};
    \node[cell, font=\scriptsize] at (3.9, 0) {SHA3-N};
    \node[cell, font=\scriptsize] at (5.2, 0) {Generic-1};
    \node[cell, font=\scriptsize] at (6.5, 0) {Generic-M};

    \foreach \x in {0,...,5} {
        \foreach \y in {1,2,3} {
            \pgfmathtruncatemacro{\isActive}{(\x<4 && \y==1) ? 1 : 0}
            \ifnum\isActive=1
                \node[active] at (\x*1.3, -\y*0.6) {};
            \else
                \node[cell] at (\x*1.3, -\y*0.6) {...};
            \fi
        }
    }

    \draw[->, line width=0.8pt] (-0.5, -0.4) -- (-0.5, 2.7);
    \draw[->, line width=0.8pt] (0.8, -0.4) -- (0.8, 2.7);
    \draw[->, line width=0.8pt] (2.1, -0.4) -- (2.1, 2.7);
    \draw[->, line width=0.8pt] (3.4, -0.4) -- (3.4, 2.7);

    \node[trace_label] at (-1.1, -0.9) {Main Trace Table};

    \foreach \x in {0,1,2,3} {
        \fill (\x*1.3 - 0.5, -0.4) circle (2pt);
    }

    \node[caption_style] at (3, -2.5) {(a) Monolithic};
\end{scope}

\begin{scope}[shift={(-2.6,-5.7)}]
    \node[constraint, minimum width=2cm] at (-0.2, 1.43) {\footnotesize Low Degree SHA3\\Constraint-1};
    \node[constraint, minimum width=2cm] at (2.55, 1.43) {\footnotesize Low Degree SHA3\\Constraint-2};
    \node[constraint, minimum width=2cm] at (5.3, 1.43) {\footnotesize Low Degree SHA3\\Constraint-3};

    \node[cell, font=\scriptsize] at (0, 0) {SHA3-1};
    \node[cell, font=\scriptsize] at (1.3, 0) {SHA3-2};
    \node[cell, font=\scriptsize] at (2.6, 0) {...};
    \node[cell, font=\scriptsize] at (3.9, 0) {SHA3-N};
    \node[cell, font=\scriptsize] at (5.2, 0) {Generic-1};
    \node[cell, font=\scriptsize] at (6.5, 0) {Generic-M};

    \foreach \x in {0,...,5} {
        \foreach \y in {1,2,3} {
            \pgfmathtruncatemacro{\isActive}{
                ((\y==1 && \x<=1) ||
                 (\y==2 && \x>=1 && \x<=2) ||
                 (\y==3 && (\x==0 || \x==3))) ? 1 : 0
            }
            \ifnum\isActive=1
                \node[active] at (\x*1.3, -\y*0.6) {};
            \else
                \node[cell] at (\x*1.3, -\y*0.6) {...};
            \fi
        }
    }

    \draw[->, line width=0.8pt] (-0.5, -0.7) -- (-0.5, 1.08);
    \draw[->, line width=0.8pt] (0.8, -0.7) -- (0.8, 1.08);

    \draw[->, line width=0.8pt] (1.75, -1.3) -- (1.75, 1.08);
    \draw[->, line width=0.8pt] (2.75, -1.3) -- (2.75, 1.08);

    \draw[->, line width=0.8pt] (0, -1.9) -- (0, -2.3) -- (5.8, -2.3) -- (5.8, 1.08);
    \draw[->, line width=0.8pt] (3.9, -1.9) -- (4.7, -1.9) -- (4.7, 1.08);

    \node[trace_label] at (-1.1, -1.0) {Main Trace Table};

    \fill (-0.5, -0.7) circle (2pt);  
    \fill (0.8, -0.7) circle (2pt);  
    \fill (1.75, -1.3) circle (2pt);  
    \fill (2.75, -1.3) circle (2pt);  
    \fill (0, -1.9) circle (2pt);
    \fill (3.9, -1.9) circle (2pt);

    \node[caption_style] at (3, -2.8) {(b) Decomposed, No Subcircuit};
\end{scope}

\begin{scope}[shift={(7,0)}]
    \node[constraint, minimum width=2.5cm] at (0.65, 2.8) {SHA3 Lookup\\Argument};

    \node[cell, font=\scriptsize] at (0, 0) {SHA3-In};
    \node[cell, font=\scriptsize] at (1.3, 0) {SHA3-Out};
    \node[cell, font=\scriptsize] at (2.6, 0) {Generic-1};
    \node[cell, font=\scriptsize] at (3.9, 0) {Generic-M};

    \foreach \x in {0,...,3} {
        \foreach \y in {1,2,3} {
            \pgfmathtruncatemacro{\isActive}{(\x<2) ? 1 : 0}
            \ifnum\isActive=1
                \node[active] at (\x*1.3, -\y*0.6) {};
            \else
                \node[cell] at (\x*1.3, -\y*0.6) {...};
            \fi
        }
    }

    \draw[dashed, ->, line width=0.8pt] (0.5, -0.7) -- (0.5, 2.3);
    \draw[dashed, ->, line width=0.8pt] (0.75, -0.7) -- (0.75, 2.3);

    \fill (0.5, -0.7) circle (2pt);
    \fill (0.75, -0.7) circle (2pt);

    \node[trace_label] at (-1.1, -0.9) {Main Trace Table};

    \begin{scope}[shift={(0,-3.2)}]
        \node[cell, font=\scriptsize] at (0, 0) {SHA3-In};
        \node[cell, font=\scriptsize] at (1.3, 0) {SHA3-Out};
        \node[cell, font=\scriptsize] at (2.6, 0) {SHA3-1};
        \node[cell, font=\scriptsize] at (3.9, 0) {SHA3-2};
        \node[cell, font=\scriptsize] at (5.2, 0) {...};
        \node[cell, font=\scriptsize] at (6.5, 0) {SHA3-N};

        \foreach \x in {0,...,5} {
            \foreach \y in {1,2,3,4,5} {
                \pgfmathtruncatemacro{\isActive}{
                    ((\y==1 && \x==0) ||
                     (\y==2 && (\x==2 || \x==3)) ||
                     (\y==3 && (\x==3 || \x==4)) ||
                     (\y==4 && \x==5) ||
                     (\y==5 && \x==1)) ? 1 : 0
                }
                \ifnum\isActive=1
                    \node[active] at (\x*1.3, -\y*0.6) {};
                \else
                    \node[cell] at (\x*1.3, -\y*0.6) {...};
                \fi
            }
        }

        \fill (0, -0.8) circle (2pt);
        \fill (1.3, -3.1) circle (2pt);

        \node[font=\small] at (3.25, 0.5) {SHA3 Auxiliary Table};
    \end{scope}

    \draw[dashed, ->, line width=0.8pt] (0, -4) -- (-1.5, -4) -- (-1.5, 1.7) -- (-0.2, 1.7) -- (-0.2, 2.3);
    \draw[dashed, ->, line width=0.8pt] (1.3, -6.3) -- (0.65, -6.3) -- (-1.7, -6.3) -- (-1.7, 1.9) -- (-0.5, 1.9) -- (-0.5, 2.3);

    \node[caption_style] at (3, -7.2) {(c) Decomposed, With Auxiliary Table};
\end{scope}

\begin{scope}[shift={(3.3,-9.7)}]
    \node[font=\normalsize] at (-6.5, 0) {\textbf{Legend:}};

    \node[cell] at (-4.5, 0.05) {...};
    \node[font=\small] at (-4.5, -0.55) {Inactive Cell};

    \node[active] at (-2.8, 0.05) {...};
    \node[font=\small] at (-2.8, -0.55) {Active Cell};

    \node[constraint, minimum width=1.2cm, font=\small] at (-0.8, 0.05) {};
    \node[font=\small] at (-0.8, -0.55) {Constraint};

    \fill (0.8, 0.05) circle (2pt);
    \draw[->, line width=0.8pt] (0.8, 0.05) -- (1.6, 0.05);
    \node[font=\small] at (4.8, 0.05) {Active Column/Cell in Constraint};

    \fill (0.8, -0.35) circle (2pt);
    \draw[dashed, ->, line width=0.8pt] (0.8, -0.35) -- (1.6, -0.35);
    \node[font=\small] at (3.8, -0.35) {Lookup Argument};
\end{scope}

\end{tikzpicture}%
}%
\Description{Diagram showing architectural variants for encoding the SHA3 opcode in zkEVMs.}
\caption{\texttt{SHA3} opcode in zkEVMs. (a) Monolithic design encodes the entire SHA3 as a single high-degree constraint. (b) Decomposed design models sub-operations with separate low-degree constraints. (c) Decomposed with auxiliary table moves SHA3 logic to a dedicated table; the main trace references auxiliary computation through lookup arguments. Pattern (b) is shown as a conceptual intermediate; production zkEVMs typically adopt (c) directly.}
\label{zkevm:fig:sha3-opcode-decompose}
\end{figure*}
\subsection{Storage Tree and Hash Substitution}
\label{zkevm:subsec:storage-rewrites}

Persistent storage, accessed through the \texttt{SLOAD} and \texttt{SSTORE} opcodes, attracts a separate family of rewrites that target the storage data structures themselves rather than the layout of the constraint system. Each storage access updates the state tree of Ethereum, a hexary MPT (each branch node has up to 16 children) whose internal nodes are hashed under Keccak-256~\cite{wood2014ethereum}. A faithful in-circuit encoding therefore pays the per-Keccak overhead noted for \texttt{SHA3} in Section~\ref{zkevm:subsec:opcode-decomposition} on every node along the trie path traversed by each read or write. 

Two complementary techniques ease this cost. The first replaces Keccak with a field-native hash designed for low constraint count, such as Poseidon~\cite{grassi2021poseidon} or MiMC~\cite{albrecht2016mimc}. Because each witness cell in the PLONKish main execution trace is itself an element of $\field$, these hashes ingest inputs and emit outputs without the bit-decomposition and recomposition that Keccak forces at every input word. Their round functions use the same field arithmetic that the constraint system already supports natively. 

The second restructures the tree itself using either a binary sparse Merkle tree or a Verkle tree. A binary sparse Merkle tree~\cite{dahlberg2016efficient} replaces the MPT with a fixed-depth binary tree in which every internal node hashes its two children identically. This eliminates two sources of in-circuit branching in the MPT, namely variable path depths and three distinct node kinds (branch, extension, leaf) with their own hash layouts. Path authentication therefore reduces to a single in-circuit hashing constraint, lowering the per-step cost of proving a storage access. A Verkle tree~\cite{Kuszmaul2018Verkle} works differently. Instead of hashing two children at each internal node, it commits to all the children of a node using a polynomial commitment. A Merkle proof reveals every sibling at every level, with per-level cost growing with the branching factor. A Verkle proof, in contrast, opens one position with a constant-size argument independent of branching. This enables each internal node to have hundreds of children while the proof for any leaf stays short. This makes Verkle trees attractive for stateless settings, in which verifiers receive proofs of state access in lieu of holding the full state. 

Scroll combines both ideas in its zkTrie, a binary sparse trie hashed with Poseidon that replaces the canonical Ethereum MPT inside the prover~\cite{scroll2024whitepaper, scrolltech2024zktrie}. Polygon zkEVM follows a comparable Poseidon-keyed sparse-tree design~\cite{polygon_zkevm}, and Linea uses a binary sparse Merkle tree~\cite{linea2022, begassat2021specification} hashed with Poseidon2~\cite{consensys2026shomei}. Because this rewrite alters only how the state tree is hashed and shaped, deployed bytecode and per-opcode runtime semantics are unchanged. Bytecode and opcode fidelity are therefore preserved, keeping the zkEVM EVM-equivalent per Table~\ref{zkevm:tab:zkevm_type_comparison}. The resulting state root, however, no longer matches that of L1 Ethereum, breaking Ethereum equivalence. Adopting this rewrite by itself therefore pushes the zkEVM beyond Type 1.

\subsection{Intermediate Representation (IR)-Based Transformations}
\label{zkevm:subsec:ir}

Bytecode-faithful zkEVMs (Types 1 through 2.5) prove the unmodified execution of deployed EVM bytecode~\cite{polygon_zkevm, scroll2024whitepaper, linea2022, begassat2021specification, taiko2024whitepaper}. Their circuit certifies the EVM state transition. However, their proving pipeline contains no compiler of its own; the bytecode is decoded into opcodes, each translated row by row into constraints. The EVM machine model prioritizes execution simplicity and deterministic gas metering rather than algebraic regularity suited to cryptographic proving, which produces the per-row constraint inflation noted earlier. Though the zk rewrites introduced in Sections~\ref{zkevm:sec:rewrites:subsec:stack-and-memory} through~\ref{zkevm:subsec:storage-rewrites} attempt to address it, some features of the EVM machine model remain only partially addressed.

One partially addressed feature is dynamic operand addressing in the operand stack of EVM. The stack pointer changes at runtime, so the operands of an opcode do not sit at trace columns fixed at circuit-design time. The stack reduction of Section~\ref{zkevm:sec:rewrites:subsec:stack-and-memory} partially addresses this by replacing the 1024-slot stack with a stack table that records each operand access. A permutation argument ties the accesses of the main trace to the rows of that table. However, the problem is that the stack table itself must be committed as a separately interpolated witness, with a dedicated stack module enforcing read-after-write consistency across its sorted rows. 

Another problem is the per-row count of selector-gated constraint polynomials. Since the executed program is unknown at circuit-design time, the constraint polynomial of every one of the 140+ EVM opcode families must be replicated at every row of the trace and gated by a per-row selector. Worst-case opcodes such as \texttt{SHA3} magnify this baseline because their monolithic encoding requires wide opcode-specific witness columns that would persist across every main-trace row. The opcode decomposition of Section~\ref{zkevm:subsec:opcode-decomposition} partially addresses this by factoring such cases into auxiliary tables linked by lookup arguments, keeping their wide internal state off the main trace. However, the challenge is that selector-polynomial commitment count scales with the number of opcode groups the universal circuit must support, regardless of how many opcodes the trace actually exercises at runtime. As a walkthrough of the overhead that bytecode fidelity imposes, consider \texttt{ADD}, the simplest of EVM opcodes. Even this trivial addition demands selector machinery to identify its opcode and stack table lookups to fetch its two operands.

An Intermediate Representation (IR)-based zkEVM addresses these aforementioned challenges by interposing a compiler between the deployed source contract and the proven trace. An IR is a structured, machine-readable form of a program emitted between the source language and the final target Instruction Set Architecture (ISA)~\cite{alfred2007compilers}. An ISA is the catalogue of primitive operations a machine such as EVM accepts, together with the encoding of their operands. In an IR-based zkEVM, the compiler translates Solidity or Yul into an IR, then emits instructions for a target ISA designed for low-cost arithmetization rather than for backwards compatibility with EVM. The proven trace certifies execution on this target ISA, not on deployed EVM bytecode. 

Among the major zkEVMs, only zkSync Era~\cite{zksync2024protocol, buterin_2022_zkevm} has taken this route, with a register-based target ISA called EraVM in contrast to the stack-based EVM. A register is a small named storage cell inside the virtual machine that holds one value at a time, and EraVM has $16$ registers, each $256$ bits wide. The zkEVM compiler reuses these $16$ registers across the program by tracking when each value is live and reassigning registers once values become dead. For example, an EraVM \texttt{ADD} instruction encodes both the opcode and three register IDs in the instruction word, e.g., $\mathsf{r}_3 \leftarrow \mathsf{r}_1 + \mathsf{r}_2$. The execution trace exposes one column per register ($16$ in total). At trace row $i$, the program counter $\pcTrace{i}$ fetches the instruction word (in contrast to just the opcode byte in Types~1 through~2.5 zkEVMs), which for the encoding above is the tuple (\texttt{ADD}, $\mathsf{r}_3$, $\mathsf{r}_1$, $\mathsf{r}_2$). The constraint reads the values of columns $\mathsf{r}_1$ and $\mathsf{r}_2$ at row $i$ and writes their sum to column $\mathsf{r}_3$ at row $i+1$. In contrast, in the stack-based EVM, the \texttt{ADD} opcode encodes only the opcode byte. When EVM fetches the \texttt{ADD} opcode, the operands are taken implicitly from the top of the runtime operand stack. Finding those operand values at trace row $i$ therefore requires tracking what sits on the stack at row $i$. That tracking is what the dedicated stack module of Section~\ref{zkevm:sec:rewrites:subsec:stack-and-memory} provides. The EraVM encoding shown above, by contrast, names the operand registers in the instruction word itself, so no auxiliary state is needed to find them. This register-based IR design retires the residue of the stack module rewrite of Section~\ref{zkevm:sec:rewrites:subsec:stack-and-memory}. EraVM, however, follows the same memory-module construction as Section~\ref{zkevm:sec:rewrites:subsec:stack-and-memory} for read-after-write consistency~\cite{zksync2024circuits}. 

The target ISA of zkSync EraVM exposes around $60$ instructions, a count significantly smaller than the 140+ EVM opcodes, largely due to dropping stack-management opcodes such as the \texttt{PUSH}, \texttt{DUP}, and \texttt{SWAP} families. A register-based design names operands directly in the instruction word, as the \texttt{ADD} example above illustrates, so these stack-management opcodes have no role and disappear from the instruction set. Each instruction in the universal circuit contributes a distinct selector-gated constraint polynomial, replicated at every row of the trace. Replacing 140+ opcodes with around $60$ instructions cuts the constraint polynomial count the prover commits per row of the universal trace. IR-based zkEVMs such as zkSync Era therefore resolve the per-row selector-gated constraint polynomial count problem that bytecode-faithful zkEVMs leave open. The compiler additionally applies standard optimization passes such as function inlining and dead code elimination~\cite{alfred2007compilers}, which shorten the trace before constraints are emitted. 

The IR pipeline trades EVM compatibility for lower constraint cost. Bytecode equivalence with L1 Ethereum is broken because the IR pipeline compiles source contracts from Solidity to a non-EVM target ISA, bypassing EVM bytecode entirely. As a result, IR-based zkEVMs preserve neither bytecode nor opcode compatibility with EVM. Under the taxonomy of Table~\ref{zkevm:tab:zkevm_type_comparison}, this places them at Type~4, the source-compatible-only tier~\cite{buterin_2022_zkevm}. The trust model also widens. A bytecode-faithful zkEVM (Types~1 through~2.5) places trust in the circuit alone, since the proof attests execution of the same contract bytecode users invoke via transactions. An IR-based zkEVM additionally trusts the compiler that produced the target-ISA program from the source contract. Concretely, the proof produced by an IR-based zkEVM attests target-ISA execution rather than direct EVM bytecode execution.

Table~\ref{zkevm:tab:ir-vs-rewrites} contrasts these comprehensive IR transformations with the component-specific rewrites discussed in previous sections.

\begin{table}[t]
\centering
\renewcommand{\arraystretch}{1.3}
\footnotesize
\begin{tabularx}{\textwidth}{
|>{\raggedright\arraybackslash\columncolor{gray!15}\hyphenpenalty=10000\exhyphenpenalty=10000\relax}X
|>{\centering\arraybackslash}X
|>{\centering\arraybackslash}X
|>{\centering\arraybackslash}X
|>{\centering\arraybackslash}X
|>{\centering\arraybackslash}X
|>{\raggedright\arraybackslash}X|}
\hline
\rowcolor{gray!25}
\textbf{Rewrite Mechanism} & \textbf{Granularity} & \textbf{Affected Aspect} & \textbf{Bytecode Fidelity} & \textbf{Opcode Fidelity} & \textbf{Requires IR} & \textbf{zkEVMs} \\
\hline
Stack and Memory Modeling (Section~\ref{zkevm:sec:rewrites:subsec:stack-and-memory}) & Local & Operand and Memory Access Model & \cmark & \cmark & \xmark{} (Optional) & Polygon~\cite{polygon_zkevm}, Scroll~\cite{scroll2024whitepaper}, Linea~\cite{linea2022, begassat2021specification}, Taiko~\cite{taiko2024whitepaper} \\
\hline
Opcode Decomposition (Section~\ref{zkevm:subsec:opcode-decomposition}) & Circuit & Constraint Circuit Structure & \cmark & \cmark & \xmark{} (Independent) & Linea~\cite{linea2022, begassat2021specification}, Polygon~\cite{polygon_zkevm}, Scroll~\cite{scroll2024whitepaper}, Taiko~\cite{taiko2024whitepaper} \\
\hline
Storage Tree and Hash Substitution (Section~\ref{zkevm:subsec:storage-rewrites}) & Local & State Tree Encoding & \cmark & \cmark & \xmark{} (Independent) & Linea~\cite{linea2022, begassat2021specification}, Polygon~\cite{polygon_zkevm}, Scroll~\cite{scroll2024whitepaper} \\
\hline
IR-Based Rewrites (Section~\ref{zkevm:subsec:ir}) & Global & Full Execution Semantics & \xmark & \xmark{} & \cmark{} (Canonical) & zkSync~\cite{zksync2024protocol} \\
\hline
\end{tabularx}
\caption{Classification of zkEVM rewrite mechanisms. \textit{Transformation Granularity} refers to the scope of the rewrite: \textit{Local} affects isolated operand behavior, \textit{Circuit} targets gate-level constraint shape, and \textit{Global} rewrites the full program.
\textit{Affected Aspect} denotes the execution-layer component impacted.
\textit{Bytecode fidelity} indicates whether the rewrite, on its own, leaves faithful bytecode attestation unchanged. \textit{Opcode fidelity} indicates whether the rewrite leaves per-opcode runtime semantics unchanged. \textit{Requires IR} indicates whether the rewrite depends on an Intermediate Representation (IR) pipeline. \textit{Canonical} marks the IR pipeline itself. \textit{Optional} marks rewrites that an IR pipeline subsumes, such as the stack and memory module that a register-based ISA retires. \textit{Independent} marks rewrites that remain necessary even under an IR pipeline. Higher-level concerns such as L1 state-root equivalence are not captured by these columns. The storage rewrite (Section~\ref{zkevm:subsec:storage-rewrites}), for instance, preserves both fidelities while breaking L1 state-root equivalence and is therefore confined to Type 2 zkEVMs.
}
\Description{A comparison table of four zkEVM rewrite mechanisms (stack and memory modeling, opcode decomposition, storage tree and hash substitution, and IR-based rewrites) across six dimensions: transformation granularity (local, circuit, or global), affected execution aspect, bytecode fidelity preservation, opcode fidelity preservation, whether IR is required, and example projects. The table shows the trade-offs between different rewrite approaches and their impact on EVM compatibility.}
\label{zkevm:tab:ir-vs-rewrites}
\end{table}
\section{Recursive zkEVMs and Proof Composition Strategies}
\label{zkevm:sec:recursive}

\label{zkevm:sec:recursion:composition}
Instead of generating a single proof per batch of transactions, recursion enables zkEVMs to compress multiple batch proofs into a single succinct proof for L1 verification. Recursion extends the prover with an embedded verifier circuit, so the recursive prover is now composed of two parts. The first is the base circuit, which is the same zkEVM circuit used so far to prove individual execution traces. The second is the embedded verifier circuit, which checks one or more existing proofs~\cite{thaler2022proofs, bitansky2013recursive}. We refer to this composite construction as the recursive circuit. 

Each application of the recursive prover takes two independent inputs. The first is one or more input proofs that the embedded verifier checks. The second is an independent statement-witness pair that the base circuit checks. The output of the recursive prover is a single proof attesting both that every input proof verifies and that the witness satisfies the statement of the base circuit. This contrasts with the non-recursive proofs studied so far, which attest only the latter. Iterating this composition replaces a set of batch proofs with one final succinct proof. The rollup contract on L1 verifies a single proof at cost independent of the number of compressed batches, and the rollup operator posts proportionally fewer proof bytes to L1. Recursive zkEVMs achieve proof composition through two distinct architectures. Vertical recursion embeds verifier circuits within each recursive layer, while horizontal recursion separates the base circuit from proof aggregation.
\subsection{Vertical Recursion}
\label{zkevm:sec:recursion:vertical}
Vertical recursion compresses $r$ batch proofs into a single final proof $\zkproof_{r-1}$. For each batch $i \in \{0, \ldots, r-1\}$, the prover produces a proof $\zkproof_i$, where $\zkproof_0$ is a base proof attesting only $\circuit(\trtab_0, \statement_0) = 0$ for the base zkEVM circuit $\circuit$, and each subsequent $\zkproof_i$ for $i \geq 1$ certifies a recursive constraint system $\circuit_i^{\mathsf{rec}}$ composed of three parts. First, it enforces $\circuit(\trtab_i, \statement_i) = 0$ for the current execution trace $\trtab_i$. Second, it embeds a recursive verifier circuit $V$ checking $V(\statement_{i-1}, \zkproof_{i-1}) = \mathsf{accept}$~\cite{thaler2022proofs}. Third, it enforces a chaining constraint $\mathsf{post}(\statement_{i-1}) = \mathsf{pre}(\statement_i)$, binding the post-state of batch $i-1$ to the pre-state of batch $i$ so that consecutive statements form a coherent transition. Equation~\ref{zkevm:eq:recursion-relation-fixed} formalizes this construction at layer $i$ of the recursion, where the public statement is $\statement_i$ and the witness is $(\trtab_i, \zkproof_{i-1}, \statement_{i-1})$.

\begin{equation}
\circuit_i^{\mathsf{rec}}(\statement_i, (\trtab_i, \zkproof_{i-1}, \statement_{i-1})) := \left[
\circuit(\trtab_i, \statement_i) = 0 \wedge V(\statement_{i-1}, \zkproof_{i-1}) = \mathsf{accept} \wedge \mathsf{post}(\statement_{i-1}) = \mathsf{pre}(\statement_i)
\right]
\label{zkevm:eq:recursion-relation-fixed}
\end{equation}

The rollup contract on L1 checks only the final proof $\zkproof_{r-1}$ against its public input $\statement_{r-1}$. The immediate predecessor proof $\zkproof_{r-2}$ and its statement $\statement_{r-2}$ are both witnesses of $\circuit_{r-1}^{\mathsf{rec}}$ and are therefore not exposed to L1, which only sees $(\statement_{r-1}, \zkproof_{r-1})$. Earlier proofs $\zkproof_0, \ldots, \zkproof_{r-3}$ are buried even deeper. Each began as the output proof of its own layer, became a witness in the layer above, and was progressively nested as a witness within a witness as recursion proceeded. Likewise, the earlier statements $\statement_0, \ldots, \statement_{r-2}$ were public inputs at their own layers, but at the final layer they live inside the witness chain alongside their corresponding proofs. Consequently, only $\statement_{r-1}$ appears in the public input of $\zkproof_{r-1}$. 

Every embedded verification is performed off-chain during proof generation by the embedded verifier $V$ at its own layer. The rollup contract on L1 checks only $\zkproof_{r-1}$ and does not recurse through the layers below. By soundness of the recursive composition, accepting $\zkproof_{r-1}$ implicitly certifies every embedded verification and chaining constraint, establishing that $\statement_0, \ldots, \statement_{r-1}$ form a coherent state-transition sequence executed by $\trtab_0, \ldots, \trtab_{r-1}$~\cite{halo2019, bitansky2013recursive}. However, anchoring the recursive proof to the canonical state maintained by the rollup contract on L1 is not supported by this sliding-window formulation of Equation~\ref{zkevm:eq:recursion-relation-fixed}. The public input $\statement_{r-1}$ does not expose the starting state $\mathsf{pre}(\statement_0)$ of the recursive proof, so the rollup contract on L1 cannot directly anchor the new recursive proof to its stored canonical state.

Closing this gap requires exposing the starting state of the recursive proof to L1. One option is to carry $\statement_0$ as a public input through every recursive layer, so the final proof exposes both $\statement_0$ and $\statement_{r-1}$~\cite{kothapalli2022nova}. Since $\statement_0$ is the statement of the first batch in the chain, its pre-state $\mathsf{pre}(\statement_0)$ already encodes the starting state of the recursive proof, and the chaining constraint $\mathsf{post}(\statement_0) = \mathsf{pre}(\statement_1)$ at layer 1 of the recursion extends this anchor forward to the next batch. At every layer $i \geq 1$, $\statement_0$ remains in the public input, and the recursive verifier check has the form $V((\statement_0, \statement_{i-1}), \zkproof_{i-1})$, with $\statement_0$ taken from the public input of the current layer. This check verifies $\zkproof_{i-1}$ against the combined public input $(\statement_0, \statement_{i-1})$ used at layer $i-1$. SNARK soundness rejects any mismatch on either component, so the current $\statement_0$ must equal the $\statement_0$ used at layer $i-1$, and by induction the same $\statement_0$ is wired unchanged through every layer all the way to the final layer $r-1$. At layer $r-1$, $\statement_0$ appears in the public input alongside $\statement_{r-1}$. The purpose of carrying $\statement_0$ to the final layer is precisely to make it part of the public input checked at L1, so the rollup contract on L1 can verify $\mathsf{pre}(\statement_0)$ against its stored canonical state. This convenience comes with a soundness obligation on the prover, who must attest that the $\statement_0$ exposed at layer $r-1$ is the same $\statement_0$ used at layer 0, and therefore that the wiring has propagated $\statement_0$ correctly through every intermediate layer. When the rollup contract on L1 checks $\zkproof_{r-1}$ against $(\statement_0, \statement_{r-1})$, it transitively certifies the intermediate statements $\statement_1, \ldots, \statement_{r-2}$, anchoring the start of the recursive proof.

The vertical-specific cost is the lack of parallelism across layers. Because the circuit at layer $i$ takes the previous proof $\zkproof_{i-1}$ as a witness, layer $i$ cannot begin until layer $i-1$ finishes, giving serial depth $\Theta(r)$ in the number of layers~\cite{halo2019, plonky2}. Adding prover hardware does not shorten this chain, since only one machine performs useful work at any time. For this reason, Polygon and zkSync do not combine many batches through a single vertical chain~\cite{polygon_zkevm, zksync2024protocol}. Their short vertical steps serve only to compress or convert an individual proof, not to aggregate batches.
\subsection{Horizontal Recursion}
\label{zkevm:sec:recursion:horizontal}

Horizontal recursion addresses the lack of parallelism in vertical recursion by organizing recursive composition as a binary aggregation tree rather than a linear chain. Unlike vertical recursion, where every recursive circuit embeds both the base circuit $\circuit$ and the verifier $V$, horizontal recursion separates these two roles between the leaves and the internal nodes of the tree. The leaves run the base circuit $\circuit$ on independent execution traces, producing base proofs $\zkproof_0, \ldots, \zkproof_{r-1}$ where each $\zkproof_i$ attests $\circuit(\trtab_i, \statement_i) = 0$. Every internal node embeds only the verifier $V$ to check its two child proofs $\zkproof_a$ and $\zkproof_b$ together with the corresponding statements $\statement_a$ and $\statement_b$, and produces a single aggregated proof whose public statement $\statement_{\mathsf{out}}$ satisfies the aggregation constraint system $\circuit_{\mathsf{agg}}$ defined in Equation~\ref{zkevm:eq:horizontal-aggregation}. As shown by this equation, the $\circuit_{\mathsf{agg}}$ circuit enforces three checks. First, both child proofs verify. Second, the chaining constraint $\mathsf{post}(\statement_a) = \mathsf{pre}(\statement_b)$ holds between the two children. Third, the output statement is set to $\statement_{\mathsf{out}} = (\mathsf{pre}(\statement_a), \mathsf{post}(\statement_b))$, exposing the endpoints of the merged batch range as the public statement of the new proof. 

\begin{equation}
\begin{split}
\circuit_{\mathsf{agg}}(\statement_{\mathsf{out}}, (\statement_a, \zkproof_a, \statement_b, \zkproof_b)) := \big[ & V(\statement_a, \zkproof_a) = \mathsf{accept} \wedge V(\statement_b, \zkproof_b) = \mathsf{accept} \\
& \wedge\, \mathsf{post}(\statement_a) = \mathsf{pre}(\statement_b) \wedge \statement_{\mathsf{out}} = (\mathsf{pre}(\statement_a), \mathsf{post}(\statement_b)) \big]
\end{split}
\label{zkevm:eq:horizontal-aggregation}
\end{equation}

The third constraint is what propagates the endpoints up the tree. Because $\statement_{\mathsf{out}}$ itself carries a well-defined $\mathsf{pre}$ and $\mathsf{post}$, and the next aggregation level applies the same chaining check on $\statement_{\mathsf{out}}$ and its sibling merged statement, the endpoint propagation goes layer by layer up to the root. By induction up the tree, the public input of the root proof is $(\mathsf{pre}(\statement_0), \mathsf{post}(\statement_{r-1}))$, the endpoints of the entire recursive proof. The rollup contract on L1 reads this pair and checks $\mathsf{pre}(\statement_0)$ against its stored canonical state. This propagation plays the same role as the $\statement_0$ wiring in the sliding-window form of vertical recursion. Because the leaves and the sibling nodes at each tree depth are independent, base proofs can be generated concurrently and sibling aggregations also run in parallel~\cite{plonky2}. Only $O(\log_2 r)$ aggregation layers suffice to compress all $r$ base proofs into the root proof. 

Production deployments such as the recursive aggregation in zkSync~\cite{zksync2024circuits} and Polygon~\cite{polygon_zkevm}, and the hierarchical chunk-and-batch aggregation in Scroll~\cite{scroll2024whitepaper}, adopt this topology directly. In production, an aggregation node often verifies several child proofs at once rather than exactly two~\cite{scroll2024whitepaper}. In systems such as Polygon and zkSync, this aggregation runs in a prover-friendly STARK whose proofs are costly to verify on-chain~\cite{polygon_zkevm, zksync2024protocol}. A final, short single-input recursion then re-proves the aggregated root proof inside a constant-size pairing-based SNARK that Ethereum verifies at constant cost. This last step is vertical, not horizontal, since one circuit embeds the verifier of a single proof and emits one smaller proof. The proof shrinks because the proof system changes to a constant-size SNARK, not because more proofs are combined. This pairing of horizontal aggregation with a final SNARK wrap is the dominant production pattern among deployed zkEVMs~\cite{scroll2024whitepaper, polygon_zkevm, zksync2024protocol, linea2022}.

\section{Application-Specific Rollup and EVM Deployment Patterns}
\label{zkevm:sec:zkvms}

While zkEVMs arithmetize the EVM directly, a parallel line of work proves Ethereum workloads on top of general-purpose Zero-Knowledge Virtual Machines (zkVMs). The constraint systems of such zkVMs encode the execution of a proof-friendly native ISA rather than the EVM. Nevertheless, such zkVMs connect to an Ethereum workload through two patterns, the application-specific rollup pattern and the EVM deployment pattern. Notable examples include Cairo~\cite{goldberg2021cairo}, SP1~\cite{succinct2024sp1}, RISC Zero~\cite{risczero2023zkvm}, OpenVM~\cite{openvm2026whitepaper}, ZisK~\cite{zisk2026docs}, Airbender~\cite{zksync2026airbender}, and Pico~\cite{brevis2025pico}.

Cairo~\cite{goldberg2021cairo} serves as the substrate for the \emph{application-specific rollup pattern} to process Ethereum workloads. Programs written in Cairo compile to the native ISA of the Cairo VM. Cairo is a general-purpose proving language rather than an EVM-style smart contract language. StarkNet~\cite{starkware2021starknet} is a rollup built atop the Cairo VM. It defines its state transition as the correct execution of the StarkNet OS, which is a Cairo program. The StarkNet OS orchestrates accounts, storage, and contract dispatch in Cairo, without inheriting the EVM account model, opcode semantics, or 256-bit word size. Accordingly, StarkNet semantics diverge from the EVM state transition function. The StarkNet OS defines its own account and storage model, and the Cairo zkVM attests valid execution of its native ISA rather than the EVM.

SP1~\cite{succinct2024sp1}, RISC Zero~\cite{risczero2023zkvm}, OpenVM~\cite{openvm2026whitepaper}, ZisK~\cite{zisk2026docs}, Airbender~\cite{zksync2026airbender}, and Pico~\cite{brevis2025pico}, in contrast, serve as the substrate for the \emph{EVM deployment pattern}. In this pattern, an Ethereum execution client such as Reth~\cite{reth2023} is compiled to the native ISA of the zkVM. The zkVM runs the client as a guest program~\cite{risczero2024zeth}. Such a client implements the EVM and the surrounding state transition logic of Ethereum in a general-purpose language (Rust for Reth). Running inside the zkVM, the client reconstructs Ethereum state transitions internally while the zkVM proves only the correct execution of the compiled binary of the client on its native ISA. Zeth on RISC Zero~\cite{risczero2024zeth}, \texttt{rsp} on SP1~\cite{succinct2024rsp}, \texttt{openvm-eth} on OpenVM~\cite{axiom2026openvmeth}, \texttt{zisk-eth-client} on ZisK~\cite{polygonhermez2026ziskethclient}, Zilkworm on Airbender~\cite{erigontech2026zilkworm}, and \texttt{pico-ethproofs} on Pico~\cite{brevis2026picoethproofs} are concrete realizations of the EVM deployment pattern. This compilation differs from the IR rewrites of Section~\ref{zkevm:sec:rewrites} in where EVM semantics reside. IR rewrites translate EVM bytecode into a proof-friendly IR whose opcodes still encode EVM-level operations, keeping IR-rewritten zkEVMs at Type~4 of the taxonomy. Under the EVM deployment pattern in zkVMs, EVM semantics live in the compiled client, not in the target ISA of the zkVM.

Both deployment patterns follow the same three-stage trust model, shown in Figure~\ref{zkevm:fig:evm-deployment-trust-layers}. The compiler (stage 1) and the runtime (stage 2) are trusted, while only the constraint system (stage 3) provides cryptographic guarantees. First, the compiler translates the input program into the native ISA of the zkVM. This translation is not verifiable, so a compiler bug could produce valid native-ISA code whose behavior diverges from the intended program. Second, the runtime interprets the native ISA and generates execution traces. This execution is not verifiable either, so a runtime bug can yield an algebraically valid trace of an incorrect computation. Third, the constraint system, which is the only verifiable stage, proves trace validity against the rules of the zkVM. 

\begin{figure}[t]
\centering
{%
\ifpreprint
  \renewcommand{\scriptsize}{\fontsize{7}{8.4}\selectfont}%
  \renewcommand{\footnotesize}{\fontsize{8}{9.6}\selectfont}%
  \renewcommand{\small}{\fontsize{9}{10.8}\selectfont}%
  \renewcommand{\normalsize}{\fontsize{10}{12}\selectfont}%
\fi
\begin{tikzpicture}[
    layer/.style={
        rectangle,
        draw=black,
        line width=0.8pt,
        minimum width=3.8cm,
        minimum height=1.5cm,
        align=center,
        font=\footnotesize
    },
    arrow/.style={
        ->,
        >=stealth,
        line width=1.5pt,
        black
    },
    label/.style={
        font=\scriptsize,
        align=left
    },
    side_label/.style={
        font=\scriptsize,
        align=center
    },
    panel_title/.style={
        font=\footnotesize\bfseries,
        align=center
    },
    note/.style={
        font=\scriptsize,
        align=center,
        text width=3.8cm
    }
]

\node[panel_title] at (-2.4, 5.7) {Application-specific\\rollup pattern};
\node[panel_title] at (2.4, 5.7) {EVM deployment\\pattern};

\node[font=\scriptsize] at (-2.4, 5.15) {High-level contract (e.g., Cairo)};
\draw[arrow] (-2.4, 5.0) -- (-2.4, 4.65);
\node[font=\scriptsize] at (2.4, 5.15) {EVM execution client (e.g., Reth)};
\draw[arrow] (2.4, 5.0) -- (2.4, 4.65);

\node[layer, minimum height=2.0cm] (Llayer1) at (-2.4, 3.6) {};
\node[font=\footnotesize\bfseries] at (-2.4, 4.25) {Stage 1: Compiler};
\node[font=\scriptsize, align=center] at (-2.4, 3.4) {High-level contract $\rightarrow$ native ISA};

\node[layer, minimum height=2.0cm] (Rlayer1) at (2.4, 3.6) {};
\node[font=\footnotesize\bfseries] at (2.4, 4.25) {Stage 1: Compiler};
\node[draw=black, line width=0.5pt, rectangle, minimum width=2.8cm, minimum height=0.7cm, font=\scriptsize, fill=gray!15, align=center] at (2.4, 3.55) {Embedded EVM interpreter\\(EVM semantics live here)};
\node[font=\scriptsize] at (2.4, 2.9) {Compiles to native ISA};

\node[layer, minimum width=8.6cm] (layer2) at (0, 1.2) {
    \textbf{Stage 2: Runtime}\\
    \scriptsize Interpret native ISA, generate trace
};
\node[layer, minimum width=8.6cm] (layer3) at (0, -1.2) {
    \textbf{Stage 3: Constraint System}\\
    \scriptsize Proves native-ISA trace
};

\draw[arrow] (-2.4, 2.6) -- (-2.4, 1.95);
\draw[arrow] (2.4, 2.6) -- (2.4, 1.95);
\draw[arrow] (0, 0.45) -- (0, -0.45);
\draw[arrow] (0, -1.95) -- node[right, pos=0.5, font=\scriptsize] {Proof $\zkproof$} (0, -2.8);

\node[side_label, rotate=90] at (-5.0, 2.4) {\scriptsize Not Verifiable (Trusted)};
\node[side_label, rotate=90] at (-5.0, -1.0) {\scriptsize Verifiable};
\draw[line width=0.5pt] (-4.5, 4.6) -- (-4.7, 4.6) -- (-4.7, 0.2) -- (-4.5, 0.2);
\draw[line width=0.5pt] (-4.5, 0.0) -- (-4.7, 0.0) -- (-4.7, -1.95) -- (-4.5, -1.95);

\node[side_label, rotate=90] at (5.0, 3.6) {\scriptsize Off-Chain};
\node[side_label, rotate=90] at (5.0, 0.0) {\scriptsize On-Chain (L2)};
\draw[line width=0.5pt] (4.5, 4.6) -- (4.7, 4.6) -- (4.7, 2.6) -- (4.5, 2.6);
\draw[line width=0.5pt] (4.5, 1.95) -- (4.7, 1.95) -- (4.7, -1.95) -- (4.5, -1.95);

\end{tikzpicture}%
}%
\caption{Three-stage trust model shared by both zkVM deployment patterns. The application-specific rollup pattern (left) compiles a high-level contract such as Cairo to the native ISA; the EVM deployment pattern (right) compiles an Ethereum execution client such as Reth. Both then pass through the same trust structure, a trusted runtime (Stage 2) followed by a constraint system (Stage 3). Only Stage 3 is cryptographically verifiable; Stages 1 and 2 are trusted in both patterns.}
\Description{Trust-model diagram for two zkVM deployment patterns arranged as a merge. The two panel titles at the top read application-specific rollup pattern on the left and EVM deployment pattern on the right. A downward arrow labeled high-level contract such as Cairo feeds the left Stage 1 box. A downward arrow labeled EVM execution client such as Reth feeds the right Stage 1 box. The left Stage 1 box is a compiler that maps the high-level contract to the native ISA. The right Stage 1 box is a compiler that produces a native-ISA binary of the execution client. Inside the right Stage 1 box, an inner rectangle labeled embedded EVM interpreter marks where EVM semantics live within the client being compiled. Both Stage 1 boxes feed downward into a single shared Stage 2 runtime box. That runtime interprets the native ISA and generates a trace. The Stage 2 box feeds a single shared Stage 3 constraint system box that proves the native-ISA trace. The Stage 3 box emits a downward arrow labeled Proof pi. A vertical bracket on the far left groups Stages 1 and 2 under the label not verifiable and trusted, and groups Stage 3 under the label verifiable. A vertical bracket on the far right groups the Stage 1 compiler stage under the label off-chain. It also groups the Stage 2 runtime and Stage 3 constraint system under the label on-chain (L2).}
\label{zkevm:fig:evm-deployment-trust-layers}
\end{figure}

As shown by Figure~\ref{zkevm:fig:evm-deployment-trust-layers}, the two patterns differ in the input to the compiler and in the structure of the produced native-ISA binary. Under the application-specific rollup pattern, the compiler stage compiles a high-level contract such as Cairo. The native-ISA program then equals the compiled contract of the user, so executing it directly produces the user-intended behavior. Trust failure can come only from compiler or runtime stages. Under the EVM deployment pattern, the compiler stage compiles an Ethereum execution client such as Reth, which already implements EVM semantics. However, beyond trusting the first two stages (compilation and execution), the user must additionally trust that the EVM execution client itself faithfully implements EVM semantics. 

The trust structure of the application-specific rollup pattern matches that of a Type~4 zkEVM. In both, the user writes high-level source code, which a trusted compiler maps to a custom ISA (in the application-specific rollups) or IR (in Type~4 zkEVMs). A trusted runtime then executes the produced ISA or IR and emits a trace. The constraint system verifies trace validity against the ISA or IR semantics, not against the semantics of the original high-level source code. The compiler and runtime sit outside the verifiable zone in both cases, so a compiler bug can produce ISA or IR whose meaning diverges from what the source author wrote, and a runtime bug can produce a trace that diverges from the intended behavior of the compiled program. The trust gap therefore has the same shape. The structure differs genuinely from Types~1 through~2.5, which verify EVM bytecode execution at the constraint level. The EVM state transition function is encoded as constraints, so EVM-level violations are cryptographically impossible. Types~1 through~2.5 carry a trust gap at the Solidity-to-bytecode step (solc), while the bytecode-to-trace step is verified rather than trusted. This removes an entire layer of trust that both application-specific zkVMs and Type~4 zkEVMs carry.

Despite the trust issues with the EVM deployment pattern, such a pattern can \emph{behaviorally} achieve what a Type~1 zkEVM achieves. Embedding a faithful Ethereum execution client inside the zkVM yields a proof attesting a correct Ethereum state transition, the same conclusion a Type~1 proof establishes at the constraint level. Hence, a zkVM running a correct Ethereum client (e.g., Zeth~\cite{risczero2024zeth}) can imitate Type~1 behavior, but the imitation rests on trusting the compiler from client source to native ISA, the runtime that executes that ISA, and the embedded client itself as a faithful EVM implementation. A Type~1 zkEVM carries none of these trust layers, as it encodes the EVM state transition function directly as constraints~\cite{buterin_2022_zkevm}.

Architecturally, zkVMs following the application-specific rollup or EVM deployment patterns are not in competition with EVM at the language level. They compete with zkEVMs on arithmetization cost, and their architectural decisions revolve entirely around that. Among the prominent zkVM architectures, Cairo follows the AIR-based approach~\cite{goldberg2021cairo}, and implements finite field transition systems where computation is modeled as algebraic state transitions over field elements. Cairo is a field-native machine that operates on elements of an approximately 252-bit prime field~\cite{goldberg2021cairo}. For field-native computation, the choice avoids the operand bit-decomposition that 256-bit EVM operands force on zkEVM constraint systems (Section~\ref{zkevm:subsection:r1cs}), and generalizes the field-native substitution that zkEVMs apply only inside hashing subcircuits via Poseidon (Section~\ref{zkevm:subsec:storage-rewrites}). The memory of Cairo is write-once~\cite{goldberg2021cairo}, meaning each address is assigned a value at most once during execution. This design simplifies the read-after-write consistency machinery (Section~\ref{zkevm:sec:rewrites:subsec:stack-and-memory}) that zkEVMs must build for the mutable byte-addressable memory of EVM~\cite{wood2014ethereum}. Even so, the constraint system of Cairo binds each read to the single write at the same address. Since each address receives at most one write, there is no most-recent write to identify, removing the temporal-ordering burden that zkEVMs must carry.

RISC Zero~\cite{risczero2023zkvm}, SP1~\cite{succinct2024sp1}, OpenVM~\cite{openvm2026whitepaper}, ZisK~\cite{zisk2026docs}, Airbender~\cite{zksync2026airbender}, and Pico~\cite{brevis2025pico} pursue the opposite design strategy. Rather than building a custom proof-friendly ISA like Cairo Assembly, they build their constraint systems around the standard Reduced Instruction Set Computer (RISC)-V instruction set, which was originally designed for general-purpose computing rather than zero-knowledge proving. The motivation is ecosystem reuse. RISC-V comes with a mature compiler ecosystem (notably LLVM~\cite{lattner2004llvm}) that maps high-level languages (e.g., C, Rust) to RISC-V machine code. RISC Zero, SP1, OpenVM, ZisK, Airbender, and Pico inherit support for the languages that ecosystem targets, whereas Cairo must build its compiler stack from scratch. The trade-off is that RISC-V semantics are not field-native, so each instruction encodes to more constraints than a purpose-built ISA would. To partly mitigate this overhead, SP1 offers precompiles, which are circuit-level specializations for common operations such as hashing and signature verification.

\section{Comparative Summary: Trade-offs Across the zkEVM Spectrum}
\label{zkevm:sec:summary}
\newcolumntype{Y}{>{\raggedright\arraybackslash\hyphenpenalty=10000\exhyphenpenalty=10000\relax}X}
\newcolumntype{Z}[1]{>{\hsize=#1\hsize\linewidth=\hsize\centering\arraybackslash}X}

\makeatletter
\@ifundefined{ifpreprint}{%
  \newif\ifpreprint
  \preprintfalse
}{}
\makeatother

\begin{table}[t]
\centering
\renewcommand{\arraystretch}{2}
\scriptsize
\setlength{\tabcolsep}{3pt}
\begin{tabularx}{\textwidth}{|>{\columncolor{gray!15}\bfseries\raggedright\arraybackslash\hyphenpenalty=10000\exhyphenpenalty=10000}p{2.0cm}|Z{0.5}|Z{1.25}|Z{0.9}|Z{2.4}|Z{0.7}|Z{0.75}|Z{0.6}|Z{0.9}|}
\hline
\rowcolor{gray!25}
& \multicolumn{2}{c|}{\shortstack{Compatibility\\ and trust}}
& \multicolumn{2}{c|}{\shortstack{Execution\\ and rewrites}}
& \multicolumn{2}{c|}{\shortstack{Data\\ structures}}
& \multicolumn{2}{c|}{\shortstack{Arithmetization}} \\
\hline
\rowcolor{gray!25}
\textbf{Project} & \textbf{Type} & \textbf{Trust} & \textbf{Binding} & \textbf{Rewrites} & \textbf{State} & \textbf{Memory} & \textbf{Word} & \textbf{System} \\
\hline
Taiko~\cite{taiko2024whitepaper} & Type 1 & \multirow{4}{=}{\centering Circuit} & \multirow{2}{=}{\centering Implicit ROM} & \multirow{1}{=}{\centering Stack and Memory Modeling, \\ Opcode Decomposition} & Keccak & \multirow{5}{=}{\centering RW-table} & \multirow{5}{=}{\centering 256-bit} & \multirow{2}{=}{\centering PLONKish} \\
\hhline{|-|-|~|~|-|-|~|~|~|}
Scroll~\cite{scroll2024whitepaper} & \multirow{3}{=}{\centering Type 2} & & & \multirow{3}{=}{\centering Stack and Memory Modeling, \\ Opcode Decomposition, \\ Storage Tree and Hash Substitution} & \multirow{3}{=}{\centering Poseidon} & & & \\
\hhline{|-|~|~|-|~|~|~|~|-|}
Polygon~\cite{polygon_zkevm} & & & \multirow{2}{=}{\centering Explicit ROM} & & & & & AIR \\
\hhline{|-|~|~|~|~|~|~|~|-|}
Linea~\cite{linea2022,begassat2021specification} & & & & & & & & \multirow{2}{=}{\centering PLONKish} \\
\hhline{|-|-|-|-|-|-|~|~|~|}
zkSync~\cite{zksync2024protocol} & Type 4 & \multirow{2}{=}{\centering Circuit + Compiler} & Non-EVM IR & IR-based Rewrite & Blake2s & & & \\
\hhline{|-|-|~|-|-|-|-|-|-|}
Cairo~\cite{goldberg2021cairo} & \multirow{2}{=}{\centering zkVM} & & \multirow{2}{=}{\centering Native ISA} & \multirow{2}{=}{\centering ---} & \multirow{2}{=}{\centering ---} & Write-once & Field\newline Element & \multirow{2}{=}{\centering AIR} \\
\hhline{|-|~|-|~|~|~|-|-|~|}
RISC Zero~\cite{risczero2023zkvm}, SP1~\cite{succinct2024sp1}, OpenVM~\cite{openvm2026whitepaper}, ZisK~\cite{zisk2026docs}, Airbender~\cite{zksync2026airbender}, Pico~\cite{brevis2025pico} & & Circuit + Compiler + Execution Client & & & & RW-table & RISC-V & \\
\hline
\end{tabularx}
\caption{Comparison of zkEVM and zkVM designs, ordered from the most Ethereum-equivalent design (Type 1) to the one with the lowest constraint cost (zkVMs). The trust surface is cumulative across circuit, compiler, and execution client. The \emph{rewrites} column lists the semantic rewrites of Section~\ref{zkevm:sec:rewrites} that each system applies. The \emph{state} column gives the family of the state-tree hash, the \emph{memory} column is the transient-memory model where RW-table is a permutation-checked read-write table, realized as a sorted table by the zkEVMs, RISC Zero, and ZisK, and as an offline multiset by SP1, OpenVM, Airbender, and Pico, and the \emph{word} column gives the operand model.}
\Description{A comparison table of twelve zkEVM and zkVM systems ordered from the most Ethereum-equivalent design to the one with the lowest constraint cost, with the lower two rows lying outside the Type 1 to 4 taxonomy. Eight dimensions are grouped under four headers. Compatibility and trust covers the compatibility type and the cumulative trust surface of circuit, compiler, and client. Execution and rewrites covers the program-binding mechanism (implicit or explicit ROM, a non-EVM intermediate representation, or a native instruction set) and the semantic-rewrite class (stack and memory modeling, opcode decomposition, storage tree and hash substitution, IR-based rewrites, or none for the zkVMs). Data structures covers the family of the state-tree hash and the transient-memory model. Arithmetization covers the operand word model and the arithmetization system of each system. Taiko is Type 1; Scroll, Polygon, and Linea are Type 2 with state-tree substitution; zkSync is Type 4 with a full intermediate-representation rewrite; and Cairo, RISC Zero, SP1, OpenVM, ZisK, Airbender, and Pico are zkVMs outside the taxonomy.}
\label{zkevm:tab:zkevm_design_matrix_binary}
\end{table}

The preceding sections each isolated one design dimension and presented it as an independent choice. Table~\ref{zkevm:tab:zkevm_design_matrix_binary} reads them together, ordered by the central trade-off between EVM compatibility and constraint complexity. Faithfully encoding the semantics of Ethereum is the costliest constraint regime (Section~\ref{zkevm:sec:arithmetization}), and relaxing that fidelity is what lowers the constraint cost (Section~\ref{zkevm:sec:rewrites}). We therefore order the surveyed systems by how far each one relaxes EVM fidelity, from the most Ethereum-equivalent design (Type 1 zkEVMs) to the one with the lowest constraint cost (zkVMs). The zkVMs sit outside the Type 1--4 taxonomy, because they do not encode the EVM at all.

The \emph{trust} column is cumulative. Systems through Type 2 keep the trust surface at the circuit, since they execute the same EVM bytecode that users invoke. An Intermediate Representation (IR) pipeline adds trust in the compiler, which is why Type 4 zkSync~\cite{zksync2024protocol} and the application-specific zkVM Cairo~\cite{goldberg2021cairo} share the same circuit-and-compiler trust surface. The EVM-deployment zkVMs RISC Zero~\cite{risczero2023zkvm}, SP1~\cite{succinct2024sp1}, OpenVM~\cite{openvm2026whitepaper}, ZisK~\cite{zisk2026docs}, Airbender~\cite{zksync2026airbender}, and Pico~\cite{brevis2025pico} add a third layer, and extend the trust to the embedded execution client (Section~\ref{zkevm:sec:zkvms}). 

Every production zkEVM shares a single universal circuit (Section~\ref{zkevm:sec:evolution}). The bytecode-faithful systems bind this circuit to the executed program. This is done through a program-counter-indexed lookup into a public ROM. This binding is what ties the proof to the actually invoked bytecode. Without it, a satisfying proof attests only that some self-consistent opcode sequence ran, not the program the transaction called. The public statement commits to the on-chain deployed bytecode that the ROM must reflect. Within this regime, the \textit{binding} column of Table~\ref{zkevm:tab:zkevm_design_matrix_binary} records a finer choice. Scroll~\cite{scroll2024whitepaper} and Taiko~\cite{taiko2024whitepaper} implement an implicit ROM design. The implicit ROM stores raw bytecode and decodes it in-circuit. It hash-checks directly against on-chain bytecode but pays a prover cost to decode the raw bytecode into selectors and operands. Polygon~\cite{polygon_zkevm} and Linea~\cite{linea2022,begassat2021specification} pre-decode selectors into an explicit ROM, removing that prover cost for decoding but moving the decode out of the circuit into a build step that must itself be audited and trusted. This choice trades prover cost against an audited build step, and it does not change how EVM-compatible a system is. These two ROM variants cover only the bytecode-faithful systems through Type 2. The Type 4 zkSync and zkVMs replace the ROM entirely. zkSync~\cite{zksync2024protocol} binds the proof to EraVM, a non-EVM target instruction set reached through an IR pipeline. The trace therefore attests execution of the EraVM target rather than of EVM bytecode (Section~\ref{zkevm:subsec:ir}). The zkVMs, on the other hand, bind the proof to their native ISA. In Cairo~\cite{goldberg2021cairo}, the proof attests execution of a Cairo program in its native ISA. In the SP1~\cite{succinct2024sp1}, RISC Zero~\cite{risczero2023zkvm}, OpenVM~\cite{openvm2026whitepaper}, ZisK~\cite{zisk2026docs}, Airbender~\cite{zksync2026airbender}, and Pico~\cite{brevis2025pico} zkVMs, the proof attests execution of a RISC-V binary. None of these zkVMs carry an EVM program at the instruction level (Section~\ref{zkevm:sec:zkvms}).

The \emph{rewrites} column of Table~\ref{zkevm:tab:zkevm_design_matrix_binary} records which semantic rewrites of Section~\ref{zkevm:sec:rewrites} each system applies, and the \emph{type} column shows what each rewrite costs in EVM fidelity. Stack and memory modeling and opcode decomposition leave bytecode and opcode fidelity intact, so they do not change the type classification of the zkEVM. Scroll~\cite{scroll2024whitepaper}, Polygon~\cite{polygon_zkevm},
and Linea~\cite{linea2022,begassat2021specification} also apply storage tree and hash substitution, a field-native hash replacing Keccak, plus a restructured tree. Scroll~\cite{scroll2024whitepaper} and Polygon~\cite{polygon_zkevm} replace Keccak with Poseidon~\cite{grassi2021poseidon}, while Linea~\cite{linea2022,begassat2021specification} substitutes it with Poseidon2~\cite{consensys2026shomei}. This hash substitution breaks Ethereum state-root equivalence and is exactly what places them beyond Type 1. zkSync~\cite{zksync2024protocol} also substitutes the state-tree hash, replacing Keccak with the non-field-native Blake2s~\cite{zksyncStorageCircuit}, proven through lookup arguments. The Type 4 classification of zkSync, however, comes from the IR-based rewrite that compiles contracts to a non-EVM target instruction set, and breaks both bytecode and opcode fidelities.

The \emph{memory} column records the transient-memory model. Taiko~\cite{taiko2024whitepaper}, Scroll~\cite{scroll2024whitepaper}, Polygon~\cite{polygon_zkevm}, and Linea~\cite{linea2022,begassat2021specification} witness the mutable byte-addressable memory of EVM through a permutation-checked read-write table. This is implemented as the stack and memory modules of Section~\ref{zkevm:sec:rewrites:subsec:stack-and-memory}, which enforce read-after-write consistency over sorted accesses. The EraVM trace of zkSync~\cite{zksync2024protocol} and the RISC-V zkVMs apply the same read-after-write argument to their memory. RISC Zero~\cite{risczero2023zkvm} and ZisK~\cite{zisk2026docs} sort their accesses into a permutation-checked table like the zkEVMs, while SP1~\cite{succinct2025sp1turbo}, OpenVM~\cite{openvm2026whitepaper}, Airbender~\cite{matterlabs2026airbenderrepo}, and Pico~\cite{yang2026sok} prove the same invariant with an offline multiset check. The read-after-write argument is therefore common to the zkEVMs and the RISC-V zkVMs alike. Only Cairo~\cite{goldberg2021cairo} departs, with a write-once memory that assigns each address at most one value, retiring the sorting constraints that enforce the most-recent write.

The \emph{word} column records the operand model of each system. The bytecode-faithful zkEVMs~\cite{taiko2024whitepaper, scroll2024whitepaper, polygon_zkevm, linea2022, begassat2021specification} inherit the uniform 256-bit word of EVM by executing EVM opcodes unmodified. zkSync~\cite{zksync2024protocol} retains that 256-bit word in the EraVM registers even at Type 4 (Section~\ref{zkevm:subsec:ir}), while neither bytecode nor opcode fidelity is preserved by the IR-based rewrites. The zkVMs alone escape it. Cairo~\cite{goldberg2021cairo} is field-native, while RISC Zero~\cite{risczero2023zkvm}, SP1~\cite{succinct2024sp1}, OpenVM~\cite{openvm2026whitepaper}, ZisK~\cite{zisk2026docs}, Airbender~\cite{zksync2026airbender}, and Pico~\cite{brevis2025pico} run on RISC-V machine words. That 256-bit operand is what forces the bit-decomposition constraints which the native words avoid (Section~\ref{zkevm:sec:arithmetization}).

\section{Applications, Open Problems, and Future Directions}
\label{zkevm:sec:open-problems}
\subsection{zkEVM Applications}
\label{zkevm:subsec:applications}

\subsubsection{Verifiable Exploit Disclosure.}
zkEVMs enable proving exploits (i.e., sequences of transactions that trigger vulnerabilities) without revealing attack vectors. The zkEVM generates a succinct proof for an execution trace containing the exploit sequence. The trace captures the state transition reflecting the exploited state (e.g., drained funds). Only initial and final state roots become public while the trace table with opcodes and transactions remains private. Hiding the trace requires the zero-knowledge property, namely a zk-SNARK~\cite{groth2016size} or zk-STARK~\cite{ben2018scalable} rather than a merely succinct argument. The proof demonstrates valid EVM execution without exposing the vulnerability path. This technique proves valuable for bug bounties and zero-day disclosures where premature revelation risks front-running. Since the zkEVM encodes the same canonical EVM semantics as mainnet, a valid proof certifies more than a mathematical fact. It guarantees that the disclosed sequence would reproduce the exploit if replayed on-chain.
\subsubsection{L1 Proof-Based Validation.}
L1 proof-based validation represents a potential transformation of the consensus model of Ethereum. In this approach, zkEVM technology would be deployed at L1. Block proposers would use zkEVM instances to generate validity proofs for each block. Validator nodes would then verify these proofs instead of re-executing all transactions~\cite{buterin2021endgame}. The fundamental barrier preventing practical deployment is proving latency. Across production zkEVMs, reported proving times for a full transaction batch range from roughly $3$ minutes to nearly $39$ minutes~\cite{chaliasos2024analyzing, stephan2025crowdprove}. This far exceeds the $12$-second block interval of Ethereum, preventing real-time validation. zkEVM rollups tolerate this latency by keeping proving off the critical path. The sequencer confirms transactions and updates rollup state immediately, while validity proofs are produced later and posted to L1 once ready. This proving runs asynchronously with respect to L1, so the prover works at its own pace, delaying only final settlement and never blocking the liveness of either chain. L1 proof-based validation has no such slack. There, each block must be proved and verified synchronously within the same block interval before validators attest, so minutes-long proving would stall consensus. This goal now sits on the Ethereum roadmap as the L1 zkEVM initiative, which targets real-time proving of $99\%$ of mainnet blocks within $10$ seconds~\cite{ethereum2025realtime}.

\subsubsection{Private Decentralized Finance (DeFi) Compliance and Auditing.}
zkEVMs enable privacy-preserving regulatory compliance for DeFi transactions~\cite{werner2022sok}. A zkEVM proves EVM bytecode execution, so compliance rules can be written as ordinary Solidity and run in the same trace as the DeFi operation. This unified execution ensures atomic enforcement. A privacy-preserving Decentralized Exchange (DEX) could thus prove trades meet regulatory thresholds while combining swap and compliance logic in one proof~\cite{sahu2023zkfi}. Regulators receive compliance guarantees without accessing private trade data. As with verifiable exploit disclosure, this confidentiality relies on the zero-knowledge property that hides the witness. The surveyed zkEVMs, however, publish the transaction data to L1 for data availability, exercising only succinctness rather than the zero-knowledge property. Privacy-focused designs such as Aztec realize confidentiality, but only by abandoning EVM compatibility~\cite{williamson2018aztec}. Achieving it within a production zkEVM therefore remains unrealized, requiring architectural modifications to handle confidential inputs while maintaining public state consistency.

\subsection{Open Problems and Future Directions}

\subsubsection{Reducing Proving Latency.}
Reducing proving latency remains an active research area. The fundamental barrier lies in EVM opcode arithmetic complexity. For instance, a single \texttt{SHA3} hash expands to roughly $150$K constraints under bit-decomposition arithmetizations~\cite{belles2022circom}. Storage operations require proving state tree updates through iterative Merkle path verifications, which represent a significant component of witness generation time. Heterogeneous transaction types force circuits to allocate resources for maximum complexity even when processing simple transfers. Research approaches target two directions. First, recursive folding schemes like Nova~\cite{kothapalli2022nova} could lower this overhead by avoiding the full recursive verifier circuit at each step (Section~\ref{zkevm:sec:recursive}). Instead of verifying the previous proof, the prover folds two instances of the same constraint system into one through a random linear combination of their committed witnesses, deferring a single succinct proof to the end. Only this lightweight folding step enters the circuit, giving constant per-step recursion overhead. The surveyed zkEVMs, however, still compose proofs through full embedded verifiers rather than folding, so the integration of folding into zkEVM-scale circuits remains an open direction. Second, while recursive aggregation such as zkTree~\cite{deng2023zktree} already combines many proofs into a single root proof with constant on-chain verification cost, lowering the proving cost of each aggregation step remains open.

\subsubsection{Hardware Acceleration.}
Current zkEVM proving relies on CPUs where multi-scalar multiplication (MSM) alone consumes over $70\%$ of computation time in zero-knowledge proof systems~\cite{ma2023gzkp}, with number-theoretic transform (NTT) adding further overhead. Bottlenecks arise because CPUs struggle to parallelize these complex mathematical calculations efficiently, and memory bandwidth limitations emerge when data exceeds processor cache capacity~\cite{qiu2024msmac}. Specialized hardware offers different approaches. GPUs deliver up to $17.6\times$ acceleration over prior GPU-based provers for proof generation~\cite{ma2023gzkp}. FPGAs achieve up to $328\times$ speedup for MSM over single-core CPUs by implementing custom circuits for field arithmetic~\cite{qiu2024msmac}. ASIC designs demonstrate roughly $5\times$ end-to-end speedups, up to $10\times$ on smaller benchmarks, with up to $77.7\times$ for individual MSM kernels~\cite{zhang2021pipezk}. However, they require new chip fabrication when protocols update.
\subsubsection{Formal Verification of zkEVM Semantics.}
Current zkEVMs prove cryptographic soundness but cannot guarantee semantic correctness~\cite{peng2025automated}. The proof system verifies that mathematical constraints are satisfied, not whether those constraints correctly implement EVM behavior. For example, a \texttt{SHA3} hash implementation might satisfy all polynomial equations yet produce incorrect outputs due to encoding errors. This gap means zkEVMs could execute transactions incorrectly while still generating valid proofs~\cite{peng2025automated}. Verifying all $140$+ EVM opcodes~\cite{wood2014ethereum} with their edge cases remains a substantial challenge for current formal methods, which cover only subsets~\cite{zhong2024parameterized}.

\subsubsection{Lack of Formal Benchmarking Framework.}
Current zkEVM benchmarking relies on self-reported metrics without standardized workloads or methodologies~\cite{chaliasos2024analyzing}. Projects omit details such as hardware specifications and batch compositions needed for reproducibility~\cite{scroll2024whitepaper, polygon_zkevm, zksync2024protocol}. Architectural diversity complicates standardization since Type 1--4 zkEVMs use different constraint systems and semantic rewrites, making direct comparisons challenging~\cite{chaliasos2024analyzing}. Transaction workloads further affect results. For instance, DeFi operations and cryptographic computations trigger vastly different constraint patterns. Effective frameworks require reference workloads from mainnet data and multi-dimensional metrics including proving time, memory, and gas costs.
\subsubsection{zkEVM Interoperability.}
Direct interoperability between different zkEVMs remains unsolved, mirroring the broader challenge of blockchain interoperability~\cite{li2025towards}. Current cross-system transfers route through L1. A user first withdraws from the source zkEVM rollup to Ethereum, which requires the batch proof of that system to be verified and finalized on L1. The user then redeposits the assets into the destination zkEVM rollup through its own bridge. Each leg pays separate L1 gas and waits for independent proof settlement, so a single transfer inherits the latency of two rollups. The barrier is that divergent circuit constructions and semantic rewrites give each system a distinct verifier. Each on-chain rollup verifier accepts only proofs in the format it was deployed to check, so a Type 1 and a Type 4 system cannot verify proofs produced by one another. This barrier disappears when chains are deployed from the same stack. Such chains share identical circuits and a common verifier, so a shared aggregation layer can settle transfers among them directly, without the L1 round-trip described above~\cite{polygon2024agglayer, zksync2024chains}.

\section{Conclusion}
\label{zkevm:sec:conclusion}

This survey has examined the constraint engineering techniques underlying zkEVMs through systematic analysis of production implementations and universal zkVMs. The analysis reveals a single organizing principle. The degree of Ethereum compatibility fixes the immediate constraint cost. That cost then propagates into the choice of dispatch, semantic rewrites, and recursion. This trade-off between EVM compatibility and constraint cost represents the defining characteristic that distinguishes zkEVMs from other zero-knowledge systems.

The Type 1--4 compatibility classification framework captures more than a simple trade-off spectrum. Each compatibility level establishes hard boundaries for what optimizations remain feasible. Type 1 systems, by preserving full Ethereum equivalence, must encode every heterogeneous execution behavior in their constraint systems. Type 4 systems, freed from the EVM bytecode- and opcode-level constraints, redesign execution semantics around ZK-friendly operations. A Type 1 commitment forbids any Ethereum fidelity-breaking rewrite, such as hash or state-tree substitution, and must handle every heterogeneous control flow pattern. The resulting constraint cost propagates into the downstream design decisions, from semantic rewrites to the recursion layer. Type 4 flexibility instead shifts complexity onto the compiler and its trusted intermediate-representation pipeline. Among the surveyed systems, no production system occupies Type 3, and each extreme of the spectrum has exactly one occupant (Table~\ref{zkevm:tab:zkevm_type_comparison}).

PLONKish suits EVM execution because its custom gates and lookup arguments encode heterogeneous operations compactly while maintaining manageable constraint degree. AIR excels at uniform state machines, as the zkVMs and Polygon zkEVM demonstrate.

The design dimensions we surveyed do not sit in isolation, and their costs ultimately converge on the recursion layer. Recursion is not itself constrained by compatibility type, yet it must absorb the prover costs that earlier compatibility and rewrite decisions impose. In practice these systems combine horizontal and vertical recursion rather than choosing between them. Horizontal aggregation compresses many batch proofs in only logarithmic tree depth and parallelizes across provers. A final short vertical step then rewraps the aggregated root into a constant-size proof for on-chain verification. Polygon and zkSync do not use vertical recursion to aggregate batches, since serial depth grows linearly with the number of layers.

Looking forward, zkEVM development faces two external pressures that may reshape current design patterns. Hardware acceleration, particularly FPGAs and ASICs tuned to specific proof systems, could lower proving cost enough to shift the prevailing architectural trade-offs. Protocol-level changes in Ethereum itself, such as state-root modifications or new proof-friendly opcodes, could reduce the compatibility burden that currently dominates zkEVM complexity. Every production system proving EVM execution either arithmetizes the EVM in a purpose-built constraint system or runs an EVM implementation as a guest program on a general-purpose zkVM~\cite{l2beat2026summary, ethproofs2026zkvms}. The compatibility versus constraint-cost trade-off explains both. The zkEVM landscape demonstrates that purpose-built zero-knowledge systems require fundamentally different approaches than general-purpose proof systems. The techniques we have surveyed, from arithmetization frameworks through dispatch mechanisms to recursion strategies, represent the current state of an actively evolving field.

\ifarxivbuild\else
  \section*{CRediT authorship contribution statement}
  \textbf{Yahya Hassanzadeh-Nazarabadi:} Conceptualization, Methodology, Investigation, Writing -- original draft, Writing -- review and editing. \textbf{Sanaz Taheri-Boshrooyeh:} Conceptualization, Methodology, Investigation, Writing -- original draft, Writing -- review and editing.

  \section*{Funding}
  This research did not receive any specific grant from funding agencies in the public, commercial, or not-for-profit sectors.

  \section*{Declaration of competing interest}
  The authors declare that they have no known competing financial interests or personal relationships that could have appeared to influence the work reported in this paper.

  \section*{Data availability}
  No data was used for the research described in the article.

  \section*{Declaration of generative AI and AI-assisted technologies in the manuscript preparation process}
  During the preparation of this work, the authors used Claude (Anthropic) in order to assist with revising the manuscript text, improving language and readability, and \LaTeX{} typesetting. After using this tool, the authors reviewed and edited the content as needed and take full responsibility for the content of the published article.
\fi

\ifarxivbuild
  \bibliographystyle{plain}
\else
  \bibliographystyle{elsarticle-num}
\fi
\bibliography{bibfile}

\end{document}